\documentclass[12pt,preprint]{aastex}
\singlespace
\slugcomment{submitted to Ap. J.}

\shorttitle{solar coronal heating}

\shortauthors{T. K. Suzuki }

\begin{document}

\title{On the Heating of the Solar Corona and the Acceleration of the 
Low-Speed Solar Wind by Acoustic Waves Generated in the Corona}

\author{Takeru Ken Suzuki}
\email{stakeru@th.nao.ac.jp}
\affil{Division of Theoretical Astrophysics,
       National Astronomical Observatory,
       2-21-1 Osawa, Mitaka, Tokyo, Japan 181-8588; 
       Department of Astronomy, Faculty of Science, 
       University of Tokyo, 7-3-1 Hongo, Bunkyo-ku, Tokyo, 
       Japan 113-0033}

\begin{abstract}
We investigate possibilities of solar coronal heating by acoustic waves 
generated not at the photosphere but in the corona, aiming at heating in 
the mid- to low-latitude corona where the low-speed wind is expected to 
come from. Acoustic waves of period $\tau\sim 100$s are triggered by  
chromospheric reconnection, one model of small scale 
magnetic reconnection events recently proposed by Sturrock. 
These waves having a finite 
amplitude eventually form shocks to shape sawtooth waves (N-waves), and 
directly heat the surrounding corona by dissipation of 
their wave energy.
Outward propagation of the N-waves is treated based 
on the weak shock theory,
so that the heating rate can be evaluated 
consistently with physical properties of the background coronal 
plasma without setting a dissipation length in an ad hoc manner.
We construct coronal structures from the upper chromosphere to 
the outside of 1AU for 
various inputs of the acoustic waves having a range of energy flux of
$F_{\rm w,0}=(1-20)\times 10^5$erg cm$^{-2}$ s$^{-1}$ 
and a period of $\tau=60-300$s. The heating by the N-wave dissipation 
effectively works in the inner corona and we find that the waves of 
$F_{\rm w,0}\ge 2\times 10^5$erg cm$^{-2}$s$^{-1}$ and $\tau \ge 60$s 
could maintain peak coronal temperature, $T_{\rm max} > 10^6$K. 
The model could also reproduce the density profile observed in the 
streamer region. However, due to its short dissipation length, the 
location of $T_{\rm max}$ is closer to the surface than the observation, 
and the resultant flow velocity of the solar wind is lower than the 
observed profile of the low-speed wind. 
The cooperations with other heating and acceleration sources with the 
larger dissipation length are inevitable to reproduce the real solar corona.

\end{abstract}


\section{Introduction}
The heating of the solar is been still poorly understood,  
and it is one of the most challenging but interesting questions  
to be solved in astrophysics. 
The origin of energy that heats the corona is generally believed to lurk
in the turbulent convective motions beneath the photosphere. A certain 
fraction of the kinetic energy of those turbulent motions is
carried up to the corona in the shape of non-thermal energy, 
such as magnetic or wave energy, 
and thermalization of such energy in the corona results in the 
heating of the surrounding plasma. 
Granule motions of the surface convection simply expect wave generation at 
the photospheric level. Possibilities of coronal heating by those 
waves have been investigated by many researchers (Osterbrock 1961; 
Ulmschneider 1971; McWhirter, Thonemann, \& Wilson 1975 ).
To date, among various modes of waves, although an Alfv\'{e}n wave has 
widely taken up as a convincing candidate in coronal heating and 
acceleration of the high-speed wind in the polar coronal holes through 
the ion-cyclotron damping mechanism (e.g. Cranmer, Field, \& Kohl 1999; 
Hollweg 1999), acoustic waves have not been regarded 
as a major heating source of the corona because of their dissipative 
character. 
Acoustic waves with a finite amplitude inevitably steepen their wave fronts to 
form shocks when they travel. 
They are strongly damped through upward propagation in the 
chromosphere, consequently the only 0.01\% of the initial wave energy could 
reach the corona (Stein \& Schwartz 1972; hereafter SS). A role 
of acoustic waves in solar coronal heating has been buried in oblivion for a 
long time.    

However, recent observations of dynamical structures of the solar 
corona (see Aschwanden, Poland, \& 
Rabin 2001 for recent review) highlight a role of acoustic waves generated not 
at the photosphere but in the corona. 
A large number of flares and flare-like events (e.g. Tsuneta et al.1992; 
Tsuneta 1996) have been observed by various telescopes until today  
(e.g. Nishio et al.1997; see Bastian, Benz, \& 
Gary 1998 for review) in 
a wide range of energy from $\sim 10^{24}$erg to $\sim 10^{32}$erg, and they 
are statistically argued in terms of supply of bulk energy 
to heat the global corona. 
Recent measurements of EUV frequency distribution, 
$\frac{dN}{dE}(E) \propto E^{-\alpha}$, of flare-like events gives 
a steeper power-law index ($\alpha=2.3 - 2.6$ by Krucker \& Benz 1998; 
$\alpha=2.0 - 2.6$ by Parnell \& Jupp 2000) on the lower energy side 
($E=10^{24}-10^{25}$erg). This fact makes us infer 
that small-scale flare-like events, such as nano-flares, might be sufficient 
(Hudson 1991) for the required energy budget 
($\gtrsim 10^{5.5}$erg cm$^{-2}$s$^{-1}$; Withbroe \& Noyes 1977). 
Several models taking into account such small-scale events  
have been introduced 
(Sturrock 1999; hereafter S99; Sturrock, Roald, \& Wolfson 2000; Roald, 
Sturrock, \& Wolfson 2000; hereafter RSW; Tarbell, Ryutova, \& Covington 1999; 
Sakai et al. 2000; Furusawa \& Sakai 2000).
Among these models, S99 proposed a model of chromospheric magnetic 
reconnection. In this model, a flux tube, newly formed as a result of 
reconnection events, oscillates vertically to excite acoustic waves in the 
corona. RSW further showed based on a simple kinetic model that this process 
possibly liberates the sufficient energy 
in the corona with the correct relevant parameters. 
A different mechanism also predicts generation of acoustic waves in the corona.
For example, spicules in the open magnetic field region effectively 
transports energy of random motions at the photosphere into the corona and 
produce longitudinal (acoustic) waves there \citep{hol92,ks99}. 
Thus, acoustic waves are supposed to be generated constantly in the corona by 
various dynamical processes. Such acoustic waves could heat the surrounding 
corona directly, unlike acoustic waves created at the photospheric level.  
They are expected to heat the inner corona effectively by the 
shock dissipation. Therefore, density at the coronal base might be large 
enough to creat dense wind flow observed as the streamer in the mid- to 
low-latitude region, 
which is believed to be connected to the low-speed solar wind \citep{hwf97}. 
Acoustic waves might become one of relevant processes working in the region  
generating the low-speed wind of which mechanisms have been poorly elucidated.

In the above models introducing production of acoustic waves in the corona, 
the authors concentrated on the amount of kinetic (wave) energy released in 
the corona. 
However, heating of coronal gas is accomplished by thermalization of 
wave energy through dissipation of such acoustic waves.
Therefore, an appropriate treatment of propagation and dissipation of these 
waves is indispensable in understanding a problem of coronal heating. 
In this paper, we employ a formulation of a weak-shock theory 
constructed by SS, originally developed to 
study acoustic waves generated by convective motions at the photosphere, 
for propagation and dissipation of the waves produced in the corona. Then, 
we can explicitly determine the rate of heating by dissipation of the waves 
on given wave energy flux (or amplitude) and wave 
period without tuning other free parameters;
we do not have to arbitrarily take an exponential type of heating of 
mechanical energy input, $F_{\rm m} \propto 
\exp(\frac{-(r-R_{\odot})}{l_{\rm m}})$, for an assumed constant 
dissipation length, $l_{\rm m}$, which is poorly 
supported by fundamental physical processes, although this conventional shape 
of the heating law was adopted, for lack of an alternative
in most previous models \citep{ko76,wtb88,sl94}, to study the global 
coronal structure.

The paper is organized as follows:
In \S\ref{sc:mdl}, we present our model. 
We briefly summarize the generation of acoustic waves in the corona
(\S\ref{sc:cmr}), and the weak-shock theory formulated by SS (\S\ref{sc:wst}). 
Basic equations describing the global coronal structure from $1 R_{\odot}$ 
to $\gtrsim$ 1AU are shown in \S\ref{sc:be}, and we present our 
method to construct a unique coronal solution on given wave parameters.  
In \S\ref{sc:rslt} we show our model results. 
First, we discuss variation of coronal wind structures with respect to  
input parameters from a theoretical point of view (\S\ref{sc:awdp}). 
Second, we examine several characteristic properties of the resultant 
corona (\S\ref{sc:rscp}). 
Finally, we test whether our model is able to reproduce recent 
observed results for the low-speed wind in the mid- to low-latitude region 
(\S\ref{sc:obs}). In \S\ref{sc:smds}, we summarize our results and discuss 
related topics.

\section{Model}
\label{sc:mdl}
\subsection{Generation of Acoustic Waves in the Corona}
\label{sc:cmr}
As a new relevant process of coronal heating in the quiet-Sun region, 
S99 proposed a model of network-field magnetic reconnection at the
chromospheric level. 
The chromosphere, specially at 
the minimum-temperature location, is quite a favorable site for reconnection 
to occur, since the magnetic resistivity is greatest, and the width of the 
current sheet, which is possibly scaled with the pressure 
scale height, is smallest there. Such reconnection events form a new closed 
magnetic flux tube, 
which is initially located far from the 
equilibrium state, and hence, will spring upward quickly (consult S99 
for a schematic picture). 
The tube would oscillate about the equilibrium state normal to the 
solar surface with a period of  
$\tau = 2L/v_{\rm A}$, where $v_{\rm A}$ is Alfv\'{e}n 
velocity
and $L$ is the  length of the tube. If we use the quoted values in 
S99 as a typical example of the tube, $L \sim 1.5\times 10^9$cm, 
a mean magnetic field, $\sim 100$G, and a mean density, 
$\sim 10^{-12}{\rm g\; cm^{-3}}$, the 
oscillation periods would be $\tau \sim 100$s. It could keep  
oscillating several tens of times before it is damped (S99). As a 
result, these perpendicular oscillations would excite longitudinal waves 
traveling in the vertical direction. Given that 
the configurations of the magnetic fields are perpendicular above the flux 
tube, such waves would propagate upwardly as acoustic waves in the 
rarefied atmosphere. They must contribute to the heating of the 
surrounding corona directly by the dissipation of the waves. 

Using a simple kinetic model, RSW have estimated the energy liberated 
by the above mechanism as a function of a mean 
magnetic field strength at the photospheric level. They showed that the mean 
field of $\sim 10$G could generate the required heating rate 
($\gtrsim 10^{5.5}$erg cm$^{-2}$s$^{-1}$) to explain the conditions in the  
quiet-Sun region. A sizable fraction of the released 
energy is supposed to be transported to the energy of the acoustic waves. 
In our model, we parameterize the input energy flux of the acoustic 
waves (not the total energy liberated by the reconnection) as $F_{\rm w,0}$
(erg cm$^{-2}$s$^{-1}$). In this paper, to study the physical processes 
clearly, we focus on the role of such 
acoustic waves in the coronal heating and construct the coronal structure 
on the input $F_{\rm w,0}$ and wave period $\tau$, though some of the released 
energy would be transformed into other MHD waves depending on 
complex configurations of magnetic fields \citep{trc99,sky00}.  

We would like to remark that other mechanisms also predict production of 
longitudinal waves at the coronal height, although we have taken up the 
chromospheric reconnection model as a typical process in this paper. 
Generation of spicules in an open magnetic flux tube have been investigated 
by various authors \citep{hjg82,hol92,ks99}. 
They found that Alfv\'{e}n waves excited by random motions at the 
photosphere \citep{ulr96} effectively transport their energy to the corona. 
The nonlinear effect of torsional Alfv\'{e}n waves 
produces longitudinal waves along the vertical flux tube in the corona. 
A sizable fraction of the initial energy of the transverse waves 
at the photosphere is 
converted to energy of the longitudinal waves at the coronal height, 
and these waves could become acoustic waves propagating upwardly. 
They could heat the surrounding plasma in the very same way as those 
triggered by the reconnection events above. 
Thus, acoustic waves are expected to be universally 
generated by various mechanisms in the corona far above the photosphere, and 
therefore, it is quite worth studying their role in coronal heating.

\subsection{Dissipation of Acoustic Waves in Corona}
\label{sc:wst}
In this section, we describe our method of treating the outward propagation 
of acoustic waves, after they have been excited in the corona. 
We first estimate a distance acoustic waves travel before forming shocks 
on a plane-parallel geometry. 
Then, we derive an equation dealing with variation of N-wave amplitude on 
spherical geometry.
For simplicity's sake, in the following discussions, we neglect the 
effects of magnetic field on the propagation of the waves. This 
simplification is valid when the 
circumstantial magnetic configuration is perpendicular   
above the flux tube generating the waves.

\subsubsection{A Distance Acoustic Waves Travel Before Forming Shocks}  
Any acoustic wave having a finite amplitude inevitably changes its shape, makes
the wave front steepen and eventually forms the shock front (e.g. Landau 
\& Lifshitz 1959).  
The distance the acoustic waves travel before forming the shocks can be 
estimated on a given wave length, $\lambda$, and initial amplitude, 
$\delta v_0$. Consider the acoustic waves propagating in the upward
direction, $z$, in isothermal atmosphere with density 
structure of $\rho = \rho_0 \exp(-z/H_{\rho})$. 
Assuming waves having initially sinusoidal 
velocity profiles, $\delta v=\delta v_0 \sin(2\pi Z/\lambda)$, 
the wave crest overtakes the preceding trough 
to form a shock front at
\begin{equation}
\label{eq:shfm}
z-z_0=2H_{\rho} \ln(1+\frac{1}{4(\gamma +1)}\frac{\lambda}{H_{\rho}}
\frac{c_{\rm s}}{\delta v_0})\; 
\end{equation} 
(SS), where $z_0$ is the position at which the waves are created, 
$\gamma=5/3$ is a ratio of specific heat, 
and $c_{\rm s}$ is a sound velocity.

In our calculations, waves of the initial amplitude, 
$\delta v_0/c_{\rm s} = 0.1 \sim 1$ are considered (\S\ref{sc:awdp}), 
and therefore, the second term in the 
logarithm in eq.(\ref{eq:shfm}) will be bounded by an upper limit: 
\begin{equation}
\label{eq:shht}
\frac{1}{4(\gamma +1)}\frac{\lambda}{H_{\rho}}\frac{c_{\rm s}}{\delta v_0} \lesssim
\frac{\lambda}{H_p} = (\frac{{c_{\rm s}}^2}{\gamma g})^{-1}{c_{\rm s} 
\tau}\simeq 0.25(\frac{\tau}{100{\rm s}})(\frac{c_{\rm s}}{2\times 
10^7{\rm cm\:s^{-1}}})^{-1}
\end{equation}
where $g$ 
is acceleration of gravity, and we have 
used $H_{\rho}\simeq H_p$ which is satisfied in the corona where temperature 
varies slowly within a scale of $\lambda(\sim 10^9{\rm cm})$.  
Then, eq.(\ref{eq:shfm}) can be expanded to a first order as
\begin{equation}
\label{eq:shfmex}
z-z_0\simeq \frac{\lambda}{2(\gamma+1)}\frac{c_{\rm s}}{\delta v_0}. 
\end{equation}
The factor $\frac{1}{2(\gamma+1)}\frac{c_{\rm s}}{\delta v_0}$ is an order of 
unity, indicating that after traveling one wave length, the waves begin to 
dissipate energy.

\subsubsection{Variation of N-wave amplitude}
The initially sinusoidal acoustic waves are expected to be 
transported as N-waves after the formation of the shocks,
provided that they are continuously generated from the lower corona. 
These N-waves are propagated outwardly in the rarefied atmosphere and 
dissipate their wave energy 
to heat the corona. For the purpose of giving a reasonable estimate of
heating rate as a function of position, variations of amplitude, 
$\delta v_{\rm w}$, of the 
transported N-waves are treated based on the weak-shock theory, 
following the formulation presented by SS. 
In the discussions below, propagation of the wave train are considered in 
a flow tube with a cross-section of $A$, to keep the consistencies with 
a model for the global corona presented in \S\ref{sc:be}. $A$ is a function 
of $r$, a distance measured from the center of the Sun, and modeled in 
\S\ref{sc:be} with taking into account non-radial expansion of the flow tube.
From now on, all the physical quantities are expressed as functions of 
$r$ only, unless explicitly declared.

An equation for the variation of wave amplitude normalized by ambient sound 
velocity, $\alpha_{\rm w}\equiv \frac{\delta v_{\rm w}}{c_{\rm s}}$, 
can be found from SS:
\begin{equation}
\label{eq:wvdr} 
\frac{1}{\alpha_{\rm w}}\frac{d\alpha_{\rm w}}{dr}=\frac{1}{2}(-\frac{1}{p}\frac{dp}{dr}+
\frac{1}{E_{\lambda}}\frac{dE_{\lambda}}{dr} - 
\frac{1}{\lambda}\frac{d\lambda}{dr}). 
\end{equation}
where $p$ is gas pressure, and $E_{\lambda} = \frac{1}{3}\rho 
(\delta v_{\rm w})^2 \lambda = \frac{1}{3} 
\gamma p \alpha_{\rm w}^2\lambda$ is a wave energy per wave length, $\lambda$.
The variation of the wave energy can be estimated from entropy generation
by the weak shock (SS; \S56 in Mihalas \& Mihalas 1984) as
\begin{equation}
\nabla\cdot{E_{\lambda}}=\frac{dE_{\lambda}}{dr}+\frac{E_{\lambda}}{A}
\frac{dA}{dr}=-E_{\lambda}\frac{2(\gamma +1) \alpha_{\rm w}}{\lambda}\simeq 
-E_{\lambda}\frac{2(\gamma +1) \alpha_{\rm w}}{c_{\rm s}\tau} ,
\end{equation}
where we have used a relation of $\lambda =c_{\rm s}(1+\frac{\gamma+1}{2}
\alpha_{\rm w})\tau \simeq c_{\rm s}\tau$ by assuming $ \alpha_{\rm w} <1$.
In general, the period of waves traveling in different 
media remains as a constant, implying
$
\frac{1}{\lambda}\frac{d\lambda}{dr}=\frac{1}{c_{\rm s}}\frac{dc_{\rm s}}{dr}.
$
Hence, eq.(\ref{eq:wvdr}) is reduced to 
\begin{equation}
\label{eq:wvdrrd} 
\frac{d\alpha_{\rm w}}{dr}=\frac{\alpha_{\rm w}}{2}(-\frac{1}{p}
\frac{dp}{dr}-\frac{2(\gamma +1) \alpha_{\rm w}}{c_{\rm s}\tau} 
-\frac{1}{A}\frac{dA}{dr} -\frac{1}{c_{\rm s}}\frac{dc_{\rm s}}{dr}). 
\end{equation}
This equation determines the variation of the wave amplitude in the 
solar corona according to the physical properties of the background coronal
plasma.  
The first term in the r.h.s. is positive in the density decreasing atmosphere, 
and the second term of the entropy generation (heating ) is negative. 
In the $(1-2)R_{\odot}$ region, where the dissipation is important to  
heat the corona, these two terms always dominate the other terms, 
the third term arising from the geometrical expansion and the fourth due to 
temperature variation.
Using $\alpha_{\rm w}$ determined above as well as background physical 
quantities, $\rho$ and $c_{\rm s}$,  
wave energy flux, $F_{\rm w} ({\rm erg\; cm^{-2} s^{-1}})$, is derived as
\begin{equation}
\label{eq:wvfx}
F_{\rm w} = \frac{1}{3}\rho c_{\rm s}^3(\alpha_{\rm w})^2 (1+ 
\frac{\gamma+1}{2}\alpha_{\rm w}), 
\end{equation}
if recalling that the wave crest moves at a speed of 
$c_{\rm s} (1 + \frac{\gamma+1}{2}\alpha_{\rm w})$.

Since the model described above is simple, we instead have several limitations
that should be taken into account with great care. 
The first obvious limitation is that the wave amplitude should 
satisfy the assumption of weak-shock, $\alpha_{\rm w}<1$.   
Second, it does not take into account the effects of gravity. 
Third, it is constructed in a static medium, therefore we cannot apply it 
to cases of moving media, such as N-waves in the solar wind where the flow 
velocity, $v$, exceeds the ambient $c_{\rm s}$.
Of the three limitations, the third one is most easily overcome,
because in all the cases we calculate in this paper, at least 99\% of the 
initial wave energy dissipates within $r<1.3 R_{\odot}$, the region where
$v\ll c_{\rm s}$ is fulfilled (figs.\ref{fig:taudp} \& \ref{fig:fwdp}). 
The second limitation also seems to have little effect if the wave period is 
small enough, since SS found that the weak-shock theory gives reasonable 
estimates for waves of $\tau < \frac{1}{2}\frac{2\pi}{\omega_{\rm ac}}$, 
where $\omega_{\rm ac}=\frac{\gamma g}{2 c_{\rm s}}$ is 
the acoustic cut-off frequency in a gravitationally stratified atmosphere,
after comparing results of the weak-shock theory with 
those of fully non-linear calculations. As 
$\omega_{\rm ac}\simeq \frac{1}{800({\rm s})}(\frac{c_{\rm s}}
{2\times 10^7{\rm cm\: s^{-1}}})^{-1}$ in the corona, the weak-shock theory 
seems to be applicable to waves of $\tau<2000$s.
To check whether the first limitation is overcome, we need to 
calculate a $\alpha_{\rm w}$ variation
in the solar corona at first hand. Our results show that any waves 
we calculate give $\alpha_{\rm w} <  0.5$ in the entire 
corona (\S\ref{sc:awdp}).  Thus, the weak-shock theory appears to be 
applicable to our model and to give a reasonable estimate of the 
heating rate as a function of $r$.

\subsection{Basic Equations}
\label{sc:be}
We here present basic equations to describe one-component
coronal wind structure in a flow tube with a cross-section of $A$ 
under a steady state condition. 
Then, an equation of continuity becomes
\begin{equation}
\label{eq:cntn}
\rho v A = {\rm const.}
\end{equation}
An equation of momentum conservation is
\begin{equation}
\label{eq:eqm}
v \frac{dv}{dr}=-\frac{G M_{\odot}}{r^2}-\frac{1}{\rho}\frac{dp}{dr}- 
\frac{1}{\rho c_{\rm s}(1+\frac{\gamma+1}{2}\alpha_{\rm w})}
\nabla \cdot F_{\rm w}.
\end{equation}
Pressure, $p$, is related to $\rho$ and temperature, $T$, by an equation 
of state for ideal gas:
\begin{equation}
\label{eq:eqst}
p=\rho \frac{k_{\rm B}}{m_{\rm H}\mu}T ,
\end{equation}
where $m_{\rm H}$ is hydrogen mass, $k_{\rm B}$ is Boltzmann constant, and  
$\mu$ is mean atomic weight of particles in unit of $m_{\rm H}$. 
The third term in eq.(\ref{eq:eqm}) represents wave momentum deposition, 
neglecting the effects of reflection and refraction \citep{rv77}.
Total energy equation is obtained as  
\begin{equation}
\label{eq:egcns}
\nabla \cdot[\rho v (\frac{1}{2} v^2 + \frac{\gamma}{\gamma-1}
\frac{k_{\rm B}}{m_{\rm H}\mu}T - 
\frac{G M_{\odot}}{r}) + F_{\rm w} + F_{\rm c}] + q_{\rm R} = 0,
\end{equation}
where we adopt the classical form of conductive flux for ionized gas, 
\begin{equation}
\label{eq:cdflx}
F_{\rm c}=-\kappa \frac{dT}{dr} = -\kappa_0 T^{5\over{2}}\frac{dT}{dr}\; ,
\end{equation}
with $\kappa_0=1.0\times 10^{-6}$ in c.g.s unit \citep{aln73}, and 
the radiative cooling term, $q_{\rm R}$(erg cm$^{-3}$s$^{-1}$), is derived from
tabulated radiative loss function, 
$\Lambda({\rm erg\: cm^3 s^{-1}})$, \citep{LM90} for the optically thin 
plasma as
$q_{\rm R}=n_{\rm e} n_{\rm p}\Lambda,$ where  $n_{\rm e}$ \& 
$n_{\rm p}$ are electron and H$^{+}$ density respectively. They are 
calculated by solving ionization 
of H and He (we ignore heavy elements) under LTE condition. 
The term, $\nabla \cdot F_{\rm w} (\le 0)$, in eq.(\ref{eq:egcns}) 
indicates the heating by the dissipation of the waves, which is explicitly 
written as 
\begin{equation}
\label{eq:wvgr}
\nabla \cdot F_{\rm w}=\cases{ 0 & $r\le r_{\rm d}$ \cr \rho v [\frac{1}{3}
\alpha_{\rm w}^2 (1+\frac{\gamma+1}{2} \alpha_{\rm w}) 
\frac{d}{dr}(\frac{c_{\rm s}^3}{v})+(\frac{2}{3}\alpha_{\rm w}
+\frac{\gamma+1}{2}\alpha_{\rm w}^2)(\frac{c_{\rm s}^3}{v})
\frac{d\alpha_{\rm w}}{dr}]& $r> r_{\rm d}$ \cr}\; ,
\end{equation}
where $r_{\rm d}$ is the position where the waves start to dissipate by 
shaping N-waves, and the height 
$h_{\rm d}\equiv r_{\rm d}-R_{\odot}$ from the photosphere 
corresponds to the sum of the height of the magnetic flux tube ($\sim 10^9$cm; 
the generation point of the waves) and the distance the acoustic waves travel 
before forming the shock 
fronts ($\sim \lambda 
\sim 10^9$cm; eq.(\ref{eq:shfmex})). 
We adopt $h_{\rm d}=2\times 10^9$cm as a standard value.

To take into account non-radial 
expansion of the flow tube due to configurations of magnetic field, 
cross-sectional area, $A$, is modeled as 
\begin{equation}
\label{eq:ftg}
A=r^2 \frac{f_{\rm max}e^{(r-r_1)/\sigma}+f_1}{e^{(r-r_1)/\sigma}+1}
\end{equation}
where
$$
f_1=1-(f_{\rm max}-1)e^{(1-r_1)/\sigma}
$$
\citep{ko76,wtb88}. 
The cross section expands from unity to $f_{\rm max}$ most drastically between 
$r=r_1-\sigma$ and $r_1+\sigma$. 
Of three input parameters, $f_{\rm max}$ 
is the most important in determining solar wind structure. In this paper, 
we consider cases for $f_{\rm max}=5$ and $f_{\rm max}=1$ (pure 
radial expansion). As for the rests of the two 
parameters, we employ $r_1=1.25R_{\odot}$ and $\sigma=0.1R_{\odot}$, 
the same values adopted in a model for 'quiet corona' in \citet{wtb88}.

\subsection{Boundary Conditions and Computational Method}
\label{sc:bc}
Now we would like to explain the practical aspects of our method of 
constructing 
wind structure with respect to various input properties of the waves.
In order to solve both heating of corona (energy transfer) and formation of
solar wind (momentum transfer) consistently, our calculation is performed 
in a broad region from an inner boundary of the 
upper chromosphere where temperature, $T_{\rm ch}=10^4$K, at 
$r_{\rm ch}=R_{\odot}+h_{\rm ch}$, which is 
located at $h_{\rm ch}=2\times 10^8$cm (2000 km) above the photosphere 
\citep{aln73}
to an arbitrary outer boundary at $r_{\rm out}=300 R_{\odot}$.

We set four boundary conditions to construct a unique solution for a given
input wave energy flux $F_{{\rm w}, 0}$ as follows:
\begin{equation}
\label{eq:bc1}
F_{\rm w}(r_{\rm ch})=F_{\rm w}(r_{\rm d})=F_{{\rm w}, 0},
\end{equation}
\begin{equation}
\label{eq:bc2}
T(r_{\rm ch})=T_{\rm ch},
\end{equation}
\begin{equation}
\label{eq:bc3}
|F_{\rm c}(r_{\rm ch})|(\simeq 0) \ll |F_{\rm c, max}|, 
\end{equation}
\begin{equation}
\label{eq:bc4}
\nabla \cdot F_{\rm c}(r_{\rm out})=0,
\end{equation}
where $F_{\rm c, max}$ in eq.(\ref{eq:bc3}) is a maximum value of downward 
conductive flux in the inner corona.
The first condition denotes that wave energy flux  
must agree with a given value when the waves start to dissipate.  
The second condition is also straightforward: the 
temperature has to coincide with the 
fixed value at the inner boundary. 
The third condition is the requirement that the downward thermal 
conductive flux should become sufficiently small at the upper chromosphere 
($T=10^4$K), diminishing from its enormous value at the coronal base 
($T\sim 10^6$K). 
Practically, we continue calculations iteratively until 
$F_{\rm c}(r_{\rm ch})/F_{\rm c, max}<1\%$ is satisfied.
The fourth condition corresponds to an ordinary requirement that no heat is
conducted inward from infinity \citep{sl94}. 
Note that thanks to the third condition, coronal base density, which is poorly 
determined from the observations, does not have to be used as a boundary 
condition. As a result, the number of free parameters to be set in advance is
reduced \citep{hm82a,hm82b,wtb88}. 
The density at the coronal base or the transition region (TR) 
is calculated as an output; 
larger input $F_{{\rm w}, 0}$ increases downward $F_{\rm c}$ in 
the lower corona, demanding larger density in the 
coronal base and TR to enhance radiative cooling to balance 
with the increased conductive heating. 

For numerical integration of the momentum equation (eq.(\ref{eq:eqm})) 
and the energy equation (eq.(\ref{eq:egcns})), we respectively use $v$ and 
an isothermal sound velocity, $a$, defined as 
\begin{equation}
a^2=\frac{c_{\rm s}^2}{\gamma}=\frac{p}{\rho}=\frac{k_{\rm B}}{m_{\rm H}\mu}T.
\end{equation}
To carry out the integration, the equations shown in the previous
section need to be 
transformed into useful forms. First, an expression for velocity gradient
can be written from eqs.(\ref{eq:cntn}),(\ref{eq:eqm}), \& (\ref{eq:eqst}): 
\begin{equation}
\label{eq:vlgr}
\frac{dv}{dr}=\frac{-\frac{G M_{\odot}}{r^2}+\frac{a^2}{A}\frac{dA}{dr}-
\frac{da^2}{dr}}{v-\frac{a^2}{v}}
\end{equation}
Second, 
an expression for gradient of $a^2$ is derived from an integrated form of 
eq.(\ref{eq:egcns}): 
\begin{equation}
\label{eq:svgr}
\frac{d a^2}{dr}=\frac{\rho v}{\kappa_0 a^5}(\frac{k_{\rm B}}{m_{\rm H}\mu})^{
\frac{7}{2}}(\frac{v^2}{2}+\frac{\gamma}{\gamma -1}a^2 -\frac{G M_{\odot}}{r}
+\frac{F_{\rm w}}{\rho v} + \int_{r_0}^{r} dr\frac{q_{\rm R}}{\rho v}
- E_{\rm tot})\; ,
\end{equation}
where, $r_0$ is a certain reference point of integration that will be 
set later. $E_{\rm tot}$ is mathematically an integral constant, which must be 
conserved in the entire region of the calculation.

Only transonic solutions are allowed for the flow speed, $v$. The 
integration of three differential equations (\ref{eq:vlgr}),(\ref{eq:svgr}), 
and (\ref{eq:wvdrrd}) is carried out simultaneously from the sonic point, 
$r=r_{\rm s}$ to 
both outward and inward directions by the fourth-order Runge-Kutta
method, also deriving density from eq.(\ref{eq:cntn}) at each integration.
We have used variable sizes for the grid of the integration, setting the 
smaller mesh size in 
the region where physical values change rapidly. For instance, in the 
TR, we set a mesh as small as $10^3$ cm 
(0.01 km) for one grid.  
To start the integration, we have to set nine variables, 
$v_{\rm s}, (\frac{dv}{dr})_{\rm s}, 
a^2_{\rm s}, (\frac{da^2}{dr})_{\rm s}, \alpha_{\rm w,s}, 
(\frac{d  \alpha_{\rm w}}{dr})_{\rm s},\rho_{\rm s}$, 
$E_{\rm tot}$ and $r_{\rm s} $ (subscript 's' denotes the sonic point). 
We can  
determine $v_{\rm s}$, $(\frac{dv}{dr})_{\rm s}$ 
and $r_{\rm s}$ for given $a^2_{\rm s}$ and $(\frac{da^2}{dr})_{\rm s}$ by 
a condition that both the numerator and the denominator 
of eq.(\ref{eq:vlgr}) are zero at $r_{\rm s}$ \citep{pkr58}. 
$(\frac{d \alpha_{\rm w}}{dr})_{\rm s}$ is also derived 
from eq.(\ref{eq:wvdrrd}) on a given $\alpha_{\rm w,s}$. 
Setting the reference point, $r_0=r_{\rm s}$, for the integration 
of radiative cooling function, we obtain $E_{\rm tot}$ as
\begin{equation}
E_{\rm tot}=[\frac{1}{2} v^2+\frac{\gamma}{\gamma-1} a^2
-\frac{G M_{\odot}}{r}+\frac{F_{\rm w}}{\rho v}-(\frac{m_{\rm H}\mu}
{k_{\rm B}})^{\frac{7}{2}}\frac{1}{\rho v} \kappa_0 a^5
(\frac{da^2}{dr})]_{\rm s}\; ,
\end{equation}
where all the variables are evaluated at the sonic point.
Now we have four variables, $a^2_{\rm s}, (\frac{da^2}{dr})_{\rm s}, 
\alpha_{\rm w,s}$, and $\rho_{\rm s}$, remaining to be regulated by the four 
boundary conditions of eqs.(\ref{eq:bc1})--(\ref{eq:bc4}).
Concrete procedures for finding a unique solution on a given $F_{{\rm w},0 }$ 
are described below.
\begin{enumerate}
\item{One makes an initial guess for $[a^2_{\rm s}, (\frac{da^2}{dr})_{\rm s}, 
\alpha_{\rm w,s}, \rho_{\rm s}]$.}
\item{The integration is performed in the outward direction from $r_{\rm s}$.
Leaving  $a^2_{\rm s}, \alpha_{\rm w,s}, \& \rho_{\rm s}$ unchanged, 
$(\frac{da^2}{dr})_{\rm s}$ are determined by carrying out the integration 
iteratively to satisfy the outer boundary condition of eq.(\ref{eq:bc4}).}
\item{The integration is performed iteratively in the inward direction,  
improving $a^2_{\rm s}, \alpha_{\rm w,s}$, and $\rho_{\rm s}$ for the fixed 
$(\frac{da^2}{dr})_{\rm s}$ until they satisfy the three inner 
boundary conditions, eqs.(\ref{eq:bc1})-(\ref{eq:bc3}). Physically, the 
condition of the wave flux (eq.(\ref{eq:bc1})) regulates 
$\alpha_{\rm w,s}$, that of temperature (eq.(\ref{eq:bc2})) regulates 
$a^2_{\rm s}$, and that of the conductive flux (eq.(\ref{eq:bc3})) regulates 
$\rho_{\rm s}$. These relations guide improvement of the respective 
initial guesses, though they are not independently approved.
Unless the above inner boundary conditions are satisfied simultaneously, 
one returns to procedure 2, preparing a new set of 
$[a^2_{\rm s}, (\frac{da^2}{dr})_{\rm s}, \alpha_{\rm w,s}, \rho_{\rm s}]$. }
\item{One can finally find a unique solution by iterating procedures 
2 and 3.}    
\end{enumerate}

\section{Results}
\label{sc:rslt}
\subsection{Theoretical Interpretation of Resultant Coronal Structures} 
\label{sc:awdp}
\begin{figure}
\figurenum{1} 
\epsscale{0.8} 
\plotone{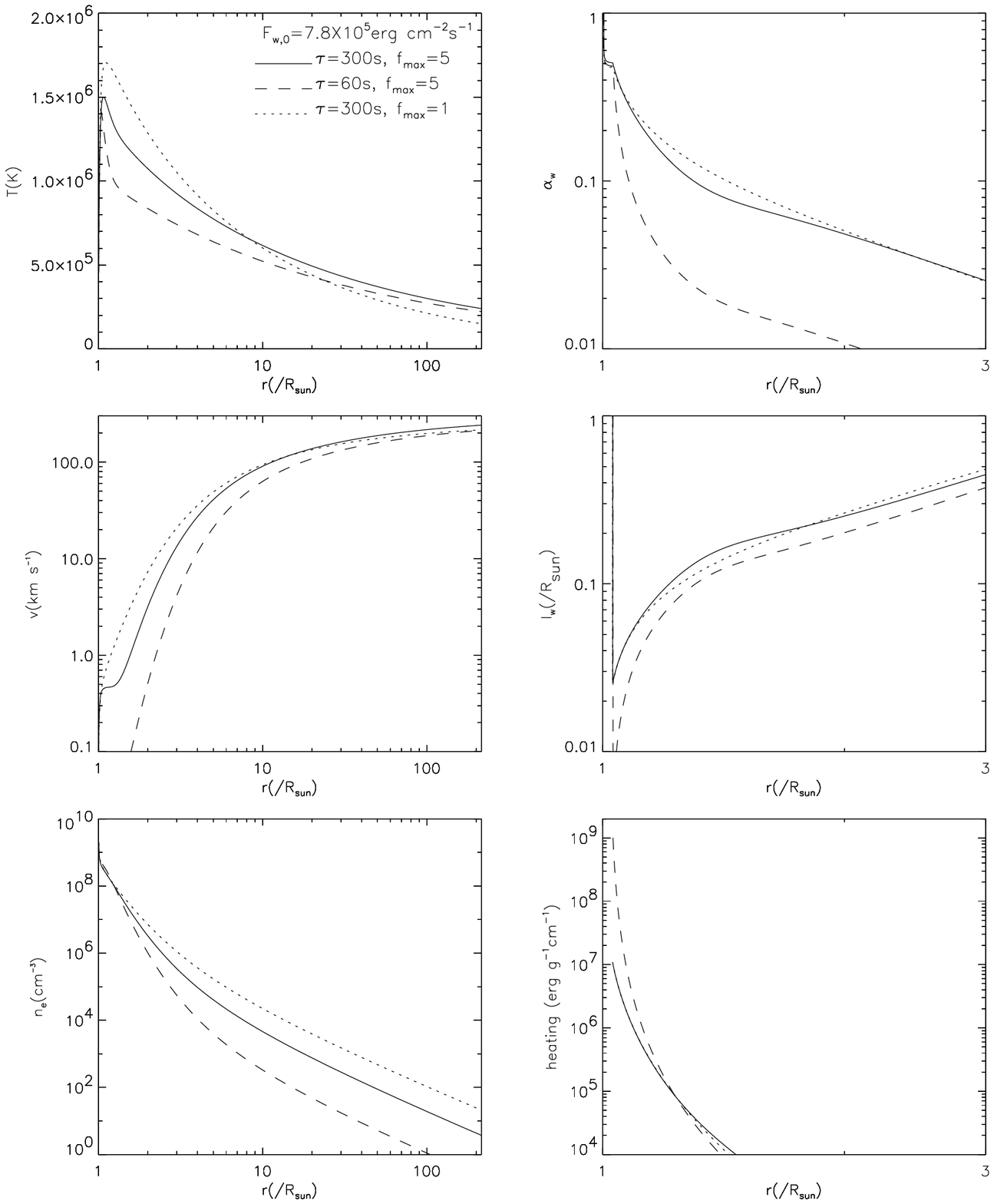} 
\caption{Variation of coronal wind structures on different sets of 
wave periods (in second) and the geometrical expansion factor, 
$(\tau,f_{\rm max})=(300,5)$ (solid), (60,5) (dashed), and (300,1) (dotted)
for the same $F_{\rm w,0}=7.8\times 10^5$erg cm$^{-2}$s$^{-1}$. 
On the left, distributions of temperature, flow velocity, and density from 
the inner boundary to 1AU ($=215 R_{\odot}$) are 
displayed from the top to the bottom. 
On the right, wave amplitude, $\alpha_{\rm w}$, scaled by ambient sound 
velocity, Dissipation length, $l_{\rm w}$, of the waves, and heating per
mass from the inner boundary to $3 R_{\odot}$ are presented 
from the top to the bottom.}
\label{fig:taudp}
\end{figure}

\begin{figure}
\figurenum{2} 
\epsscale{0.8} 
\plotone{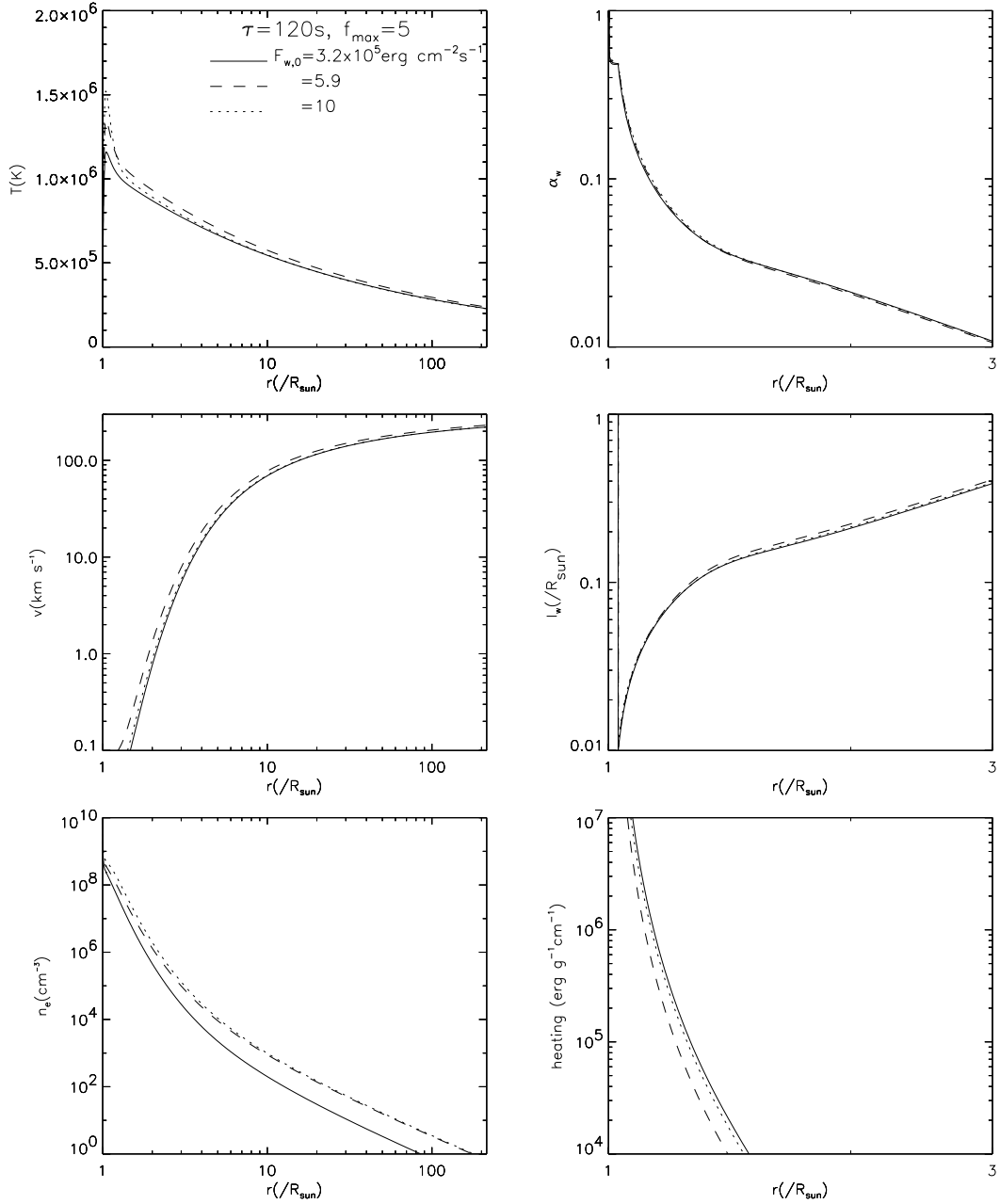} 
\caption{Variation of coronal wind structures on three different 
$F_{\rm w,0}=(3.2,5.9,10)\times 10^5$erg cm$^{-2}$s$^{-1}$ (solid, 
dashed, and dotted lines, respectively) for $\tau=120$s and $f_{\rm max}=5$.
Each panel is same as fig.\ref{fig:taudp}.}
 \label{fig:fwdp}
 \end{figure}

Figure \ref{fig:taudp} demonstrates our result of 
coronal wind structures employing the same input 
$F_{\rm w,0}=7.8\times10^5$erg cm$^{-2}$s$^{-1}$, but different sets of 
wave periods and non-radial expansion factors, 
$(\tau({\rm s}),f_{\rm max})=(300,5), (60,5), (300,1)$.  
Figure \ref{fig:fwdp} compares the wind structures adopting three different 
$F_{\rm w,0}=(3.2,5.9,10)\times10^5$erg cm$^{-2}$s$^{-1}$ for 
identical inputs of $\tau=120$s and $f_{\rm max}=5$. 
Distribution of temperature, flow velocity, and electron density from 
the inner boundary to 1AU ($=215 R_{\odot}$) are displayed from top to 
bottom on the left side of the both figures. On the right side, we present 
variation of wave amplitude, $\alpha_{\rm w}$, a dissipation length, 
$l_{\rm w}$, of N-waves, which is defined as
$l_{\rm w}=\frac{F_{\rm w}}{\nabla \cdot F_{\rm w}} $, and heating per
unit mass, $|\frac{1}{\rho v} \nabla \cdot F_{\rm w}|$, respectively. 
In tab.\ref{tab:mdr}, we tabulated several resultant properties of  
the corona and solar wind as well as the input parameters. 
\begin{table}[h]
\begin{tabular}{|l||l|l|l|l|l|}
\hline
\hline
\multicolumn{1}{|c|}{Input} & \multicolumn{5}{|c|}{Output} \\ \hline
\multicolumn{1}{|c|} {$(F_{\rm w,0},\tau,f_{\rm max})$}
& {$p_{\rm tr}$}
& {$T_{\rm max}$} & {$r_{T{\rm max}}$} 
& {$(n_{\rm p}v)_{\rm 1AU}$} & {$(v)_{\rm 1AU}$}
\\ \hline
(7.8, 60, 5) & 0.28 & 1.43 & 1.04 & $3.8\times 10^6$ & 222 \\ \hline
(7.8, 300, 5) & 0.24 & 1.50 & 1.08 & $7.5\times 10^7$ & 251 \\ \hline
(7.8, 300, 1) & 0.29 & 1.71 & 1.12 & $3.8\times 10^8$ & 222  \\ \hline
(3.2, 120, 5) & 0.13 & 1.16 & 1.05 & $2.5\times 10^6$ & 230 \\ \hline
(5.9, 120, 5) & 0.21 & 1.35 & 1.05 & $1.3\times 10^7$ & 240 \\ \hline
(10, 120, 5) & 0.32 & 1.52 & 1.05 & $1.3\times 10^7$ & 230 \\ \hline
\end{tabular}
\caption{Input parameters and output wind properties of each model; 
$F_{\rm w,0}$ is in $10^5$erg cm$^{-2}$s$^{-1}$ and $\tau$ in second. 
$p_{\rm tr}$ is pressure in dyn cm$^{-2}$ at the TR where 
$T=10^5$K, $T_{\rm max}$ is peak coronal temperature in $10^6$K, 
$r_{T {\rm max}}$ is location of 
$T_{\rm max}$ in $R_{\odot}$, $(n_{\rm p}v)_{\rm 1AU}$ is proton flux in 
cm$^{-2}$s$^{-1}$ at 1AU, and $(v)_{\rm 1AU}$ is the flow velocity in 
km/s at 1AU.} 
\label{tab:mdr}
\end{table}

To begin with, we would like 
to emphasize that corona heats up to $>10^6$K in every case, because acoustic 
waves generated in the corona are able to heat the surrounding gas directly, 
unlike acoustic waves produced at the photosphere.   
However, N-waves are rapidly damped, so that the heating occurs only in 
inner region as seen in the lower right panels of figs.\ref{fig:taudp} and 
\ref{fig:fwdp}. Then, location of maximum temperature, $T_{\rm max}$, is quite 
close to the surface. In an outside region of $\gtrsim 1.5R_{\odot}$, heat  
is input only by outward thermal conduction, and the flow is 
accelerated mostly by thermal pressure. As a result, speed of the 
solar wind 
at 1AU is $\lesssim 300$km/s, which is 
slightly slower than the actual low-speed wind ($300 \sim 450$km/s).

The top right panel of fig.\ref{fig:fwdp} 
interestingly illustrates that distributions of 
$\alpha_{\rm w}$ are almost identical in spite of very different inputs of
$F_{\rm w,0}$. Particularly, initial N-wave amplitudes, 
$\alpha_{\rm w}(r_{\rm d})$, at $r_{\rm d}$, are within a range 
between 0.48 and 0.49. 
This is because $F_{\rm w,0}$ ($\sim p \alpha_{\rm w,0}^2 c_{\rm s}$; 
eq.(\ref{eq:wvfx})) mostly owes its variation to change 
of ambient pressure (see \S \ref{sc:tmxptr}).
Moreover, an upper right panel of fig.\ref{fig:taudp} also indicates that 
initial $\alpha_{\rm w}(\simeq 0.5)$ is almost independent of $\tau$ and 
$f_{\rm max}$. 
We have found that $0.45 < \alpha_{\rm w}(r_{\rm d}) < 0.52$ within our 
parameter regions of $F_{\rm w,0}=(1-20)\times 10^5$erg cm$^{-2}$s$^{-1}$, 
$\tau=60-300$s, and $f_{\rm max}=1-5$. 
This proves that  
$\alpha_{\rm w} < 1$ is fulfilled in the entire region, which 
justifies the assumption of weak shock.
Middle right panels of figs.\ref{fig:taudp} and \ref{fig:fwdp} show that 
the dissipation length is a drastically varying function on $r$. 
This implies that the assumption of
a constant dissipation length usually taken in previous models for 
the global corona \citep{wtb88,sl94} is very poor for our N-wave process.

In the following discussions, we examine dependences of the wind structures 
on the respective input parameters. 
First, we argue dependences on wave periods. As illustrated in 
fig.\ref{fig:taudp},  
N-waves with smaller $\tau$ dissipate more quickly and the heating occurs in 
thinner region close to the surface. 
This simply leads to deposition of wave energy 
in denser region. Since radiative loss, $q_{\rm R}$(erg cm$^{-3}$s$^{-1}$), is 
in proportion to $\rho^2$ for optically thin plasma,
a greater fraction of energy supplied in denser region goes into radiative 
escape. Consequently, a smaller amount of energy remains to heat the corona 
and accelerate the flow. 
The case adopting smaller $\tau$(=60s) gives 
lower temperature in the corona, 
and therefore, a smaller pressure scale height and a more rapid 
decrease of density, as shown in fig.\ref{fig:taudp}. 
Lower temperature also takes the sonic 
point more distant from the solar surface, and then, mass flux of 
the solar wind becomes much 
smaller than that expected from the $\tau=300$s case (tab.\ref{tab:mdr}).

Second, we study effects on areal expansion of the flow tube.
Comparing results adopting the same 
$F_{\rm w,0}=7.8\times 10^5$erg cm$^{-2}$s$^{-1}$ 
and $\tau=300$s but different $f_{\rm max}=1$ and 5 in 
fig.\ref{fig:taudp}, one can notice significant change of density structure.
The model considering the non-radial expansion gives more drastic decrease 
of density as a function of $r$ in spite of similar initial density at the 
inner boundary. 
Temperature in the inner corona is also lower in that model, 
since more fraction of the input energy is lost adiabatically due to 
geometrical expansion of the flow tube. 
On the other hand, decrease of temperature is slower and 
temperature in the outer region ($\gtrsim 8 R_{\odot}$) is higher. This is 
because the lower density reduces both radiative cooling and adiabatic loss. 
The higher temperature also leads to larger acceleration of the flow there, 
giving larger speed of the solar wind in the outer corona. 

Third, we investigate dependencies on input wave energy flux.
Models employing larger input $F_{\rm w,0}$ give larger density in the inner 
corona (fig.\ref{fig:fwdp}), 
because radiative cooling should be raised to offset the increased heating. 
More accurately, larger input $F_{\rm w,0}$ results in larger downward 
conductive flux from the corona to the chromosphere, which needs larger 
radiative cooling to balance with enhanced conductive heating. 
Larger $F_{\rm w,0}$ also leads to higher temperature in the inner corona. 
However, more rapid decrease of temperature occurs because of enhanced 
radiative escape. A model employing 
$F_{\rm w,0}=10\times 10^5$erg cm$^{-2}$s$^{-1}$ 
yields lower temperature in a region of $r\gtrsim 1.5 R_{\odot}$ than that 
adopting $F_{\rm w,0}=5.9\times 10^5$erg cm$^{-2}$s$^{-1}$. 
Therefore, a larger input of $F_{\rm w,0}$ does not simply anticipate larger 
acceleration of the flow, and relations of 
$F_{\rm w,0}-(n_{\rm p}v)_{\rm 1AU}$ and $F_{\rm w,0}-(v)_{\rm 1AU}$ 
does not show simple positive correlations (tab.\ref{tab:mdr}, see also 
\S \ref{sc:npvel})

\subsection{Some Characteristic Properties of Resultant Corona} 
\label{sc:rscp}
\subsubsection{Peak Temperature and Pressure at the TR}
\label{sc:tmxptr}
\begin{figure}
\figurenum{3} 
\epsscale{1} 
\plotone{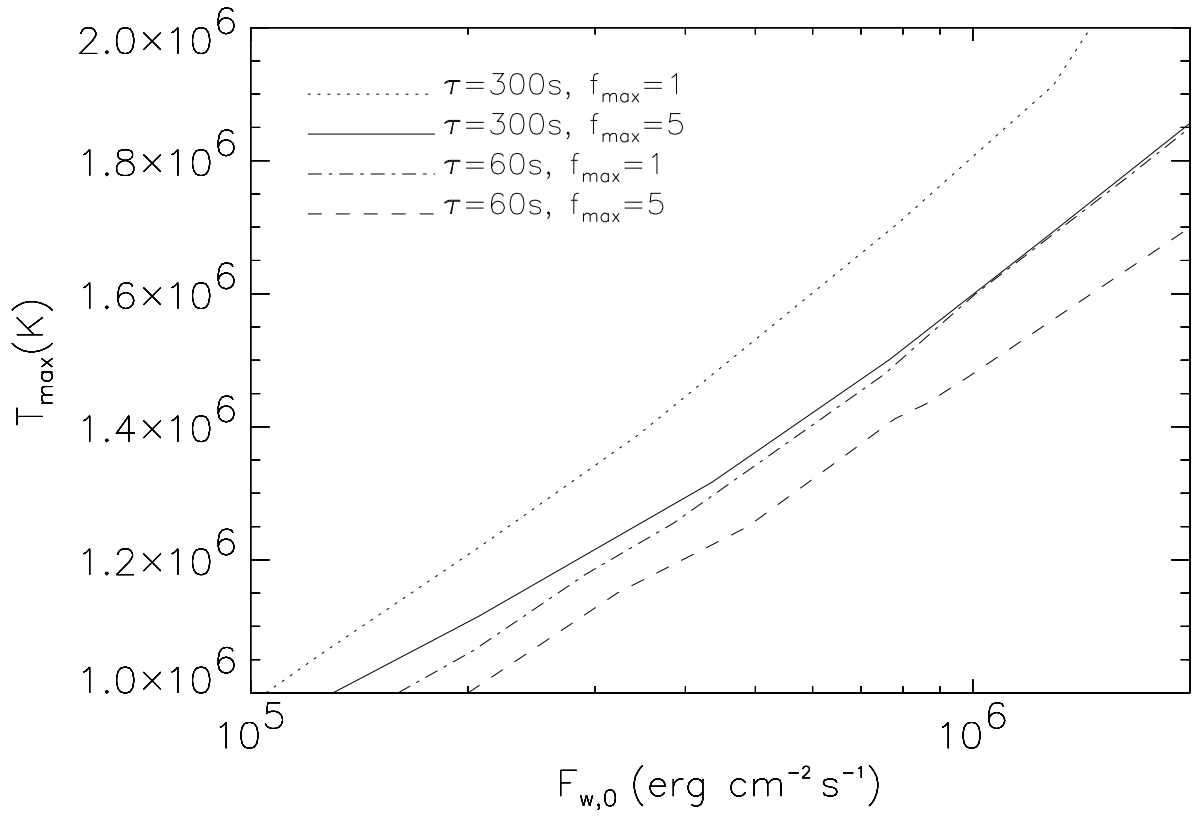} 
\caption{Relation between $F_{\rm w,0}$ and $T_{\rm max}$ on different sets 
of $(\tau,f_{\rm max})$.}
\label{fig:fwtmx}
\end{figure}

We would like to inspect relations of peak coronal temperature, $T_{\rm max}$, 
and pressure, $p_{\rm tr}$, at the TR with respect to 
the input model parameters.
In fig.\ref{fig:fwtmx}, we present $T_{\rm max}$ as a 
function of input $F_{\rm w,0}$ for different 
$\tau$ and $f_{\rm max}$. 
The figure shows that $T_{\rm max}>10^6$K is 
accomplished for $F_{\rm w,0}\ge 2\times 10^5$erg cm$^{-2}$s$^{-1}$, even if 
one chooses waves with a short period ($\tau=60$s) and a large 
expansion factor ($f_{\rm max}=5$).
$T_{\rm max}$ is a monotonically increasing function of 
$F_{\rm w,0}$ on each set of $(\tau, f_{\rm max})$, and the relations can be 
fitted by power-law as 
\begin{equation}
\label{eq:factmx}
T_{\rm max} \propto (F_{\rm w,0})^{k}, \;\;\;\;
{\rm where} \;\;\;\;\; k=0.23-0.26.
\end{equation}

\begin{figure}
\figurenum{4} 
\epsscale{1} 
\plotone{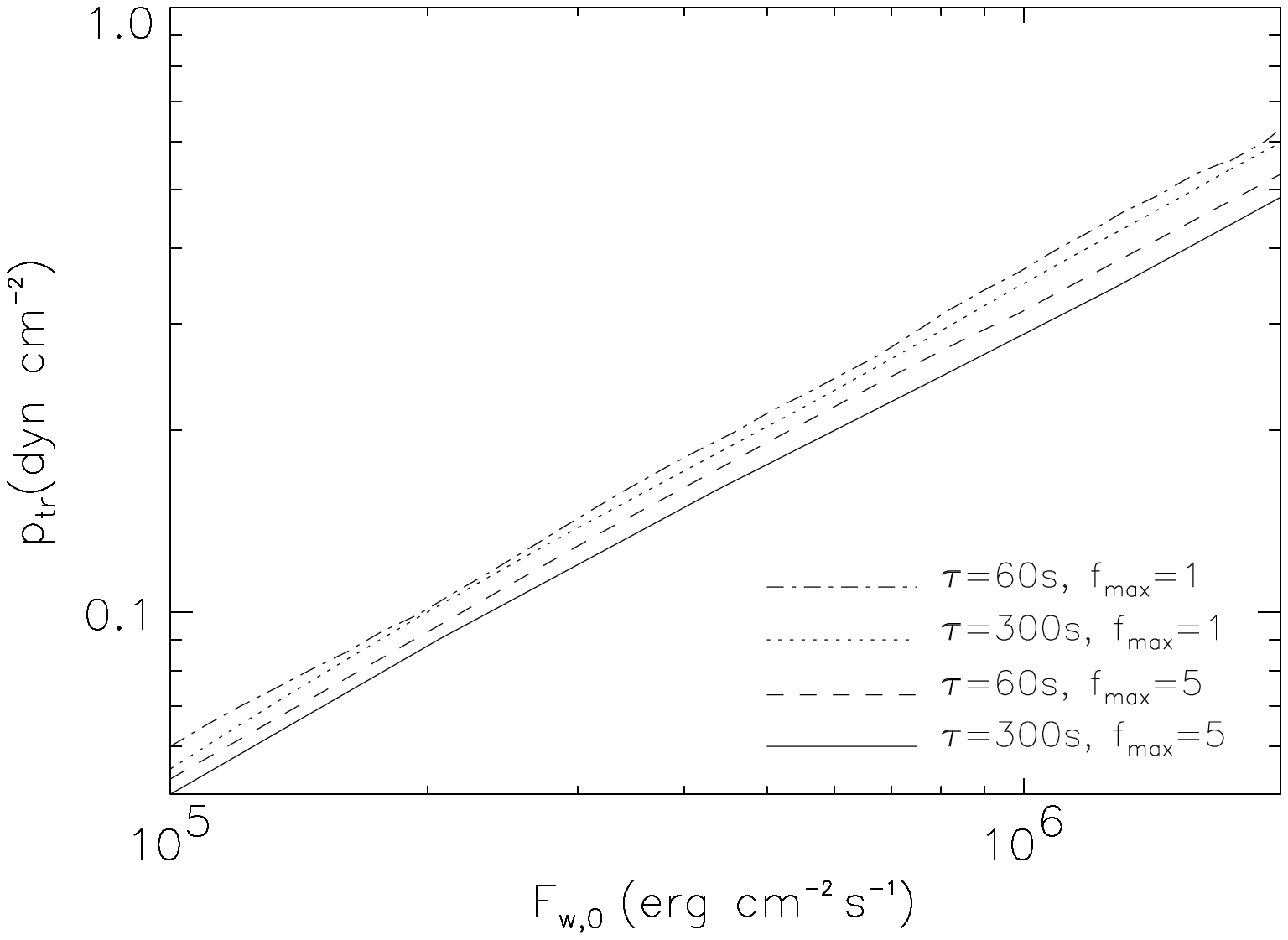} 
\caption{Relation between $F_{\rm w,0}$ and $p_{\rm tr}$ on different sets 
of $(\tau,f_{\rm max})$.}
\label{fig:fwptr}
\end{figure}

Figure \ref{fig:fwptr} displays relations between $F_{\rm w,0}$ and  
$p_{\rm tr}$. 
$p_{\rm tr}$ is evaluated at $T=10^5$K, whereas $p_{\rm tr}$ is almost a 
constant through the TR from the upper chromosphere ($T\simeq 10^4$K; inner 
boundary) to the coronal base ($T\simeq 5\times 10^5$K) because of its 
geometrically thin configuration.
The figure indicates that $p_{\rm tr}$ weakly depends on $\tau$ and 
$f_{\rm max}$, and has a positive correlation with $F_{\rm w,0}$: 
\begin{equation}
\label{eq:fwptr}
p_{\rm tr} \propto (F_{\rm w,0})^{l}, \;\;\;\;\; {\rm where} \;\;\; l=0.75-0.79
\end{equation} 
A rise of $F_{\rm w,0}$ leads directly to an increase in downward conductive 
flux, which demands higher density (or pressure) in the lower corona and the 
TR to raise radiative cooling to balance with the enhanced 
conductive heating. Alternatively,
it could be interpreted that injection of larger energy in the corona can 
heat more deeply to the higher density chromosphere by the downward 
thermal conduction.

A combination of the above two relations of 
eqs.(\ref{eq:factmx}) and (\ref{eq:fwptr}) roughly give 
$$
T_{\rm max} \propto (p_{\rm tr})^{1/3},
$$
on given $\tau$ and $f_{\rm max}$, which reminds us of the 
famous RTV scaling law \citep{rtv78}, 
$T_{\rm max} \simeq 1400 (p_{\rm l} L_{\rm h})^{1/3}$, 
for closed magnetic loops, where $p_{\rm l}$(dyn cm$^{-2}$) is loop 
pressure and $L_{\rm h}$(cm) is loop height.
Taking $r_{\rm Tmax}$ instead of $L_{\rm h}$ for our model, 
we can actually derive a relation between $T_{\rm max}$ and 
$(p_{\rm tr} r_{\rm Tmax})$ from our results as 
$$T_{\rm max} \simeq 3000 (p_{\rm tr} r_{\rm Tmax})^{0.30},$$ 
showing a form analogous to that of the original RTV law. 
This is because we consider the same energy balance among thermal conduction, 
radiative cooling, and heating with the same boundary condition that the 
conductive flux should become almost zero at the base (eq.(\ref{eq:bc3})), 
although the configurations are quite different (closed loop for the RTV law 
and open flow tube for ours). The slight discrepancies of the prefactor 
and power-law index are caused by the fact that the RTV law was 
derived on the assumption of spatially uniform heating along the loop, while 
our heating function is determined by eq.(\ref{eq:wvgr}), which is not uniform 
at all.

\subsubsection{Mass Flux and Coronal Energy Loss}
\label{sc:npvel}
\begin{figure}
\figurenum{5} 
\epsscale{1} 
\plotone{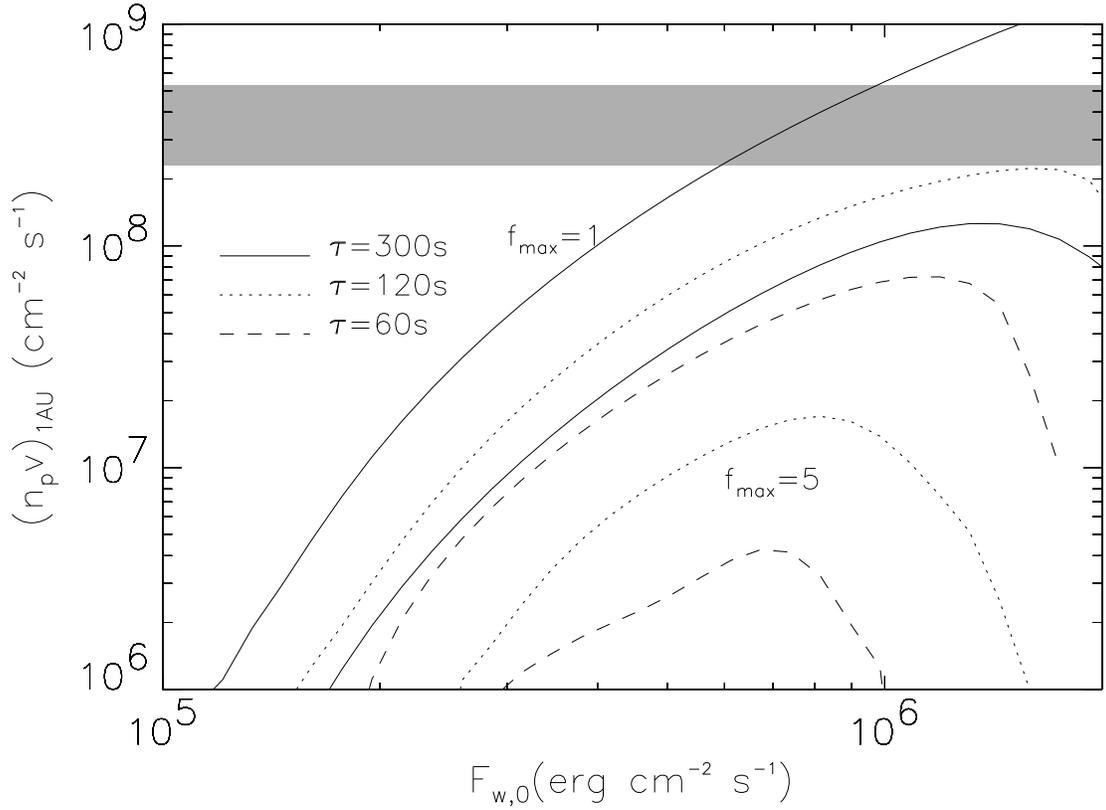} 
\caption{Relation between $F_{\rm w,0}$ and $(n_{\rm p}v)_{\rm 1AU}$ on 
$f_{\rm max}=1$ and 5. Solid lines are results for cases 
adopting $\tau=300$s, dotted lines, 120s, and dashed lines, 60s.
Shaded region is the observational constraint by \citet{wtb88} for 
'quiet corona'.}
 \label{fig:fwmd}
 \end{figure}

In fig.\ref{fig:fwmd} we show anticipated proton 
flux, $(n_{\rm p}v)_{\rm 1AU}$, at 1AU as a function of 
$F_{\rm w,0}$ for different $\tau$ and $f_{\rm max}$, with observational 
constraints, $(n_{\rm p}v)_{\rm 1AU}=(3.8\pm 1.5)\times 10^8$cm$^{-2}$s$^{-1}$ 
(Shaded), compiled by \citet{wtb88} as the empirical value for 
'quiet corona' which 
is supposed to corresponds to the mid- to low-latitude region generating 
the low-speed wind. Larger $\tau$ waves 
give a greater $(n_{\rm p}v)_{\rm 1AU}$ owing to the effective transport of 
dissipated energy to solar wind flow by avoiding radiative escape. 
Introduction of the non-radial expansion of the flow tube reduces 
$(n_{\rm p}v)_{\rm 1AU}$, 
because input energy per unit flow tube normalized at 1AU 
decreases on increasing $f_{\rm max}$, even though one inputs identical wave 
energy flux at the inner corona. Therefore, the larger areal expansion 
straightforwardly reduces mass flux of the solar wind.  
As to dependences on $F_{\rm w,0}$, $(n_{\rm p}v)_{\rm 1AU}$ has an upper 
limit for given $\tau$ and $f_{\rm max}$. 
(Even models of $\tau=300$s and $f_{\rm max}=1$ are supposed to have
the upper limit in a $F_{\rm w,0}>2\times 10^6$erg cm$^{-2}$s$^{-1}$ area.) 
Figure \ref{fig:fwmd} shows that only one case employing $(\tau,f_{\rm max})
=(300,1)$ can reproduce observed proton flux of the slow wind. 
Unfortunately more realistic cases considering the non-radial expansion 
of $f_{\rm max}=5$ cannot explain the observations, which implies that 
other mechanisms of heating (and acceleration) necessarily work cooperatively. 

\begin{figure}
\figurenum{6} 
\epsscale{1} 
\plotone{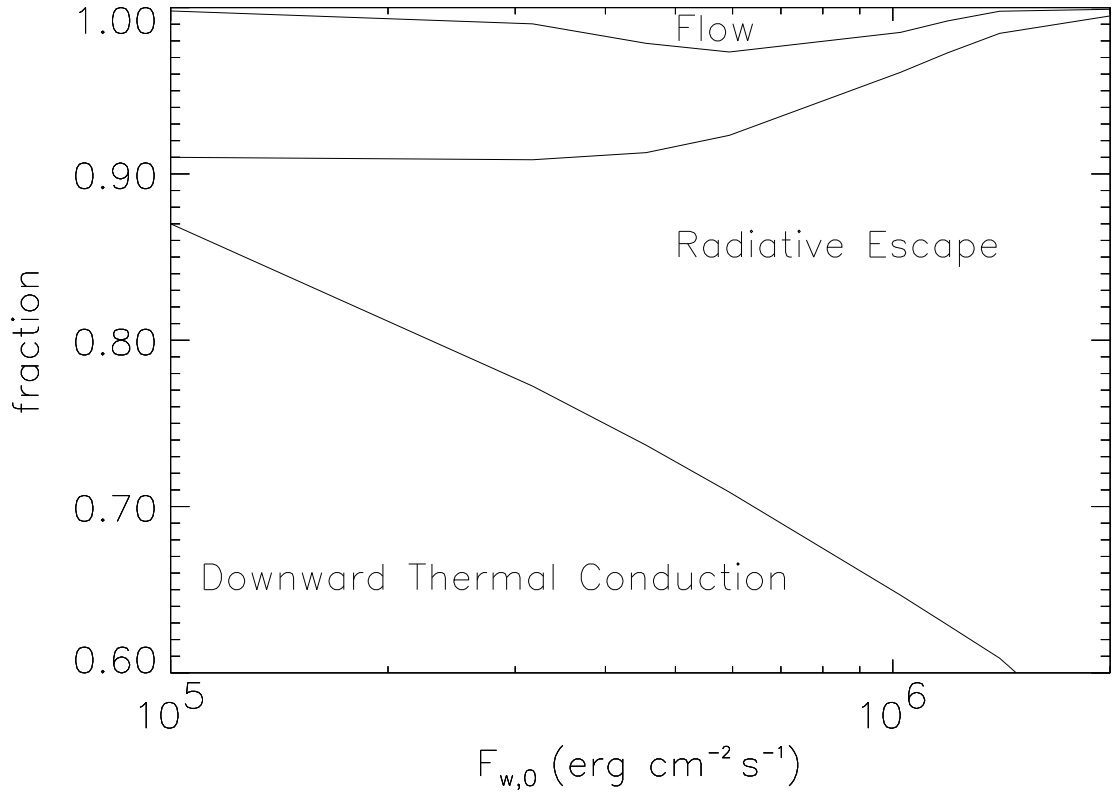} 
\caption{Fraction of three types of coronal energy loss, downward thermal 
conduction, radiative escape, and flow of the solar wind (mass loss), as a 
function of $F_{\rm w,0}$ for the $\tau=120$s waves in the flow tube with 
$f_{\rm max}=5$. 
The non-labeled region between 'radiative escape' and 
'flow' indicates energy carried away to the outside of 
$r_{\rm out}=300R_{\odot}$ by outward conduction. }
 \label{fig:fwfrc}
 \end{figure}

\begin{figure}
\figurenum{7} 
\epsscale{1} 
\plotone{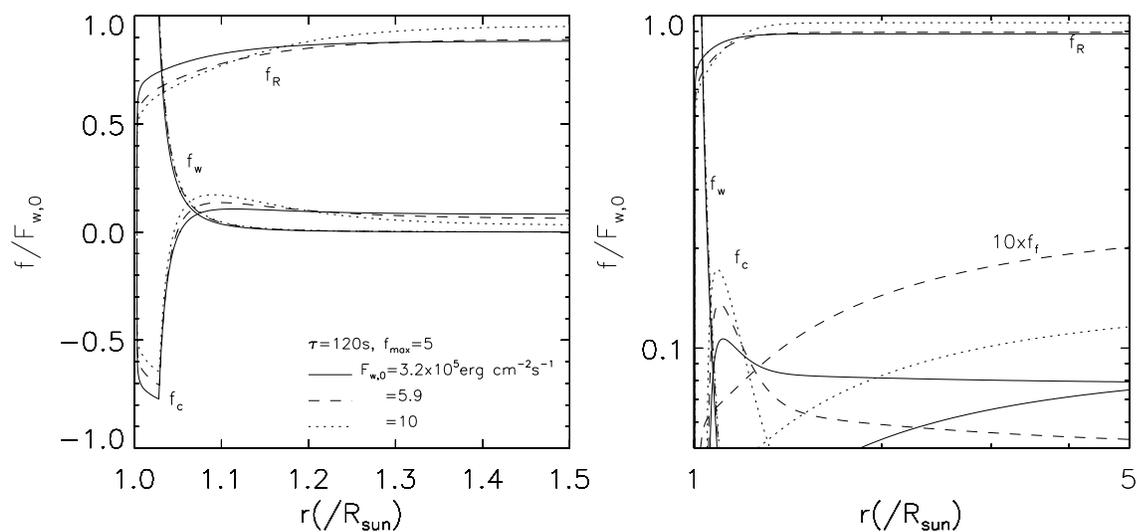} 
\caption{Exchange of energy among four components of energy flux, wave, 
radiation, conduction, and flow (consult text for detail). The 'flow' 
component is increased by 10 times. 
Models adopt the same inputs of $\tau=210$s and $f_{\rm max}=5$, but different 
$F_{\rm w,0}=(3.2,5.9,10)\times 10^5$erg cm$^{-2}$s$^{-1}$ (solid, 
dashed, and dotted lines, respectively)}
 \label{fig:egtr}
 \end{figure}

We would like to study in detail why $(n_{\rm p}v)_{\rm 1AU}$ 
has an upper limit on fixed $\tau$ and $f_{\rm max}$.
Figure \ref{fig:fwfrc} displays the ratios of the main three types of 
energy loss: 
downward thermal conduction from the coronal base, radiative escape in the 
corona, and total amount of energy converted to the flow (mass loss) 
at $r_{\rm out}$, respectively normalized by $F_{\rm w,0}$ for the cases 
adopting $\tau=120$s and $f_{\rm max}=5$. 
The values for downward thermal conduction in
fig.\ref{fig:fwfrc} are taken from maximum conductive flux $F_{\rm c, max}$ 
in lower corona (see fig.\ref{fig:egtr})
and the values for the radiative escape are
derived by the integration of the radiative cooling function from points 
with $F_{\rm c, max}$ to $r_{\rm out}$. 
The region not labeled between 'radiative escape' and 'flow' denotes
energy carried out of $r_{\rm out}$ by thermal conduction. 
According to fig.\ref{fig:fwfrc}, the main source of coronal energy 
loss is downward thermal conduction within our range of $F_{\rm w,0}$ 
(strictly speaking,  most of the conducted energy finally radiates away), 
whereas radiative escape comes to play a significant role for larger
$F_{\rm w,0}$ since density in the corona becomes higher, being subject to 
eq.(\ref{eq:fwptr}). 
Although a ratio of energy transfered to the flow is as small as
$\lesssim 3\%$ throughout the range, it has a bimodal tendency. In 
$F_{\rm w,0}<6\times 10^5$erg cm$^{-2}$s$^{-1}$ 
it increases, which implies that the larger $F_{\rm w,0}$ is, the more 
effectively the energy is transfered to the flow.  However, in 
$F_{\rm w,0}>6\times 10^5$erg cm$^{-2}$s$^{-1}$, it decreases because of the 
abrupt dominance of radiative cooling. As a result, the predicted $(n_{\rm p}
v)_{\rm 1AU}$ increases rapidly on increasing $F_{\rm w,0}$ at first and 
eventually decreases as seen in fig.\ref{fig:fwmd}.

To examine these differences in terms of energy transfer, 
we show variations of energy flux of four components, 
\begin{equation}
{\rm wave}\; :\; f_{\rm w} = \frac{A(r)}{A(r_{\rm ch})}F_{\rm w},
\end{equation}
\begin{equation}
{\rm conduction} \; : \;  f_{\rm c} = \frac{A(r)}{A(r_{\rm ch})}F_{\rm c},
\end{equation}
\begin{equation}
\label{eq:flx3}
{\rm flow}\; :\; f_{\rm f} = \rho v \frac{A(r)}{A(r_{\rm ch})}
[(\frac{1}{2} v^2+\frac{\gamma}{\gamma-1} a^2-\frac{G M_{\odot}}{r})-
(\frac{1}{2} v^2+\frac{\gamma}{\gamma-1} a^2
-\frac{G M_{\odot}}{r})_{r_{\rm ch}}],
\end{equation}
\begin{equation}
\label{eq:flx4}
{\rm radiation}\; :\; f_{\rm R} = \rho v \frac{A(r)}{A(r_{\rm ch})}
\int_{r_{\rm ch}}^{r}dr \frac{q_{\rm R}}{\rho v},
\end{equation}
per flow tube with a cross section of $A=$1cm$^2$ at the inner boundary.
The 'flow' term of eq.(\ref{eq:flx3}) contains three ingredients,
kinetic energy of the solar wind, enthalpy, and gravitational energy.  
Note that, 
except the wave energy flux, the zero point of the energy 
is taken here at the inner boundary, to clarify 
what fraction of the input energy flux is transfered to the other 
components. In a left panel of fig.\ref{fig:egtr}, the status 
of energy transfer in the inner corona is displayed in a linear scale for 
both X and Y-axis, and in a right panel, that in the broader region 
is shown in $\log$ scale. 
The ratio of the downward thermal 
conduction, $f_{\rm c}/F_{\rm w,0}$, is smaller for larger $F_{\rm w,0}$, 
though the absolute value of $F_{\rm c}$ is increasing 
along with $F_{\rm w,0}$. 
At the TR, most of the heat flux by the downward conduction finally escapes as 
radiation, except for a tiny fraction transfered to the enthalpy to be 
used to heat the TR. This leads to a smaller ratio of the radiative loss, 
$f_{\rm R}/F_{\rm w,0}$, for larger $F_{\rm w,0}$ in the inner region of 
$<1.2 R_{\odot}$. 
However, the contribution from the radiation 
continues in a much more distant region in the model employing
largest $F_{\rm w,0}$, and $f_{\rm R}/F_{\rm w,0}$ finally outdoes the 
other two cases. Consequently, the ratio of energy transfered 
to the solar wind becomes smaller than the model adopting smaller 
$F_{\rm w,0}(=5.9\times 10^5$erg cm$^{-2}$s$^{-1})$ as seen in the 
right panel. It can be concluded that 
if $F_{\rm w,0}$ is larger than a certain threshold, an increase of the input 
$F_{\rm w,0}$ does not lead to effective heating of the corona to 
accelerate the solar wind but results in deposition of the wave energy 
in the high density region to be wasted as radiative escape.

\subsubsection{Wave Amplitude in the Inner Corona}
Ultraviolet and X-ray emission lines of the corona show non-thermal 
broadenings \citep{hrsh90,edpw98}, 
which are inferred to originate from wave motions. 
We investigate whether the obtained wave amplitudes are consistent with these 
observations. 
In tab.\ref{tab:dvwm}, we show our results of the wave rms velocities in the 
inner corona, which are calculated as 
$$
<\delta_{\rm w}>=\frac{1}{\sqrt{3}}\delta_{\rm w}
$$
for the N-shaped wave.
\begin{table}[h]
\begin{tabular}{|l||l|l|}
\hline
\hline
\multicolumn{1}{|c|}{Input} & \multicolumn{2}{|c|}{$<\delta_{\rm w}>$} \\ \hline
\multicolumn{1}{|c|} {$(F_{\rm w,0},\tau,f_{\rm max})$}
& $T=5\times 10^5$K & $T=10^6$K
\\ \hline
(7.8, 60, 5) & 32 & 42 \\ \hline
(7.8, 300, 5) & 34 & 44 \\ \hline
(7.8, 300, 1) & 32 & 41  \\ \hline
(3.2, 120, 5) & 31 & 41 \\ \hline
(5.9, 120, 5) & 32 & 42 \\ \hline
(10, 120, 5) & 34 & 44 \\ \hline
\end{tabular}
\caption{$<\delta_{\rm w}>$(km/s) in the inner corona} 
\label{tab:dvwm}
\end{table}
As the waves propagate upwardly from a location of $T=5\times 10^5$K to that 
of $T=10^6$K, the wave amplitude increases according as the ambient density 
decreases. 
Our models give the very similar results for different $F_{\rm w,0}$ (also 
$\tau$ and $f_{\rm max}$), since change of $F_{\rm w,0}(\sim \rho 
<\delta_{\rm w}>^2 c_{\rm s})$ mostly reflects variation of the density 
(fig.\ref{fig:fwdp} and \S \ref{sc:awdp}). 
The tabulated results are marginally consistent with non-thermal velocities of 
$20 \sim 40$km/s obtained from observations of inner coronal lines in 
the solar disks and limbs \citep{edpw98}, 
if we assume the observed non-thermal components totally consist of the 
waves of the acoustic mode.  
However, this assumption may be too extreme, because 
other modes of the waves actually exist in the real solar corona. 

\subsection{Comparison of Coronal Wind Structure with Observed Low-Speed Wind}
\label{sc:obs}
In this section, we study a feasibility of the process of acoustic waves 
in the coronal heating by comparing our results 
with recent observations. Aiming at coronal heating and wind acceleration in a 
region where the low-speed wind is formed, we take observational data of 
the coronal streamer in the mid- to low-latitude region. 

\subsubsection{Density Distribution}
\begin{figure}
\figurenum{8} 
\epsscale{1} 
\plotone{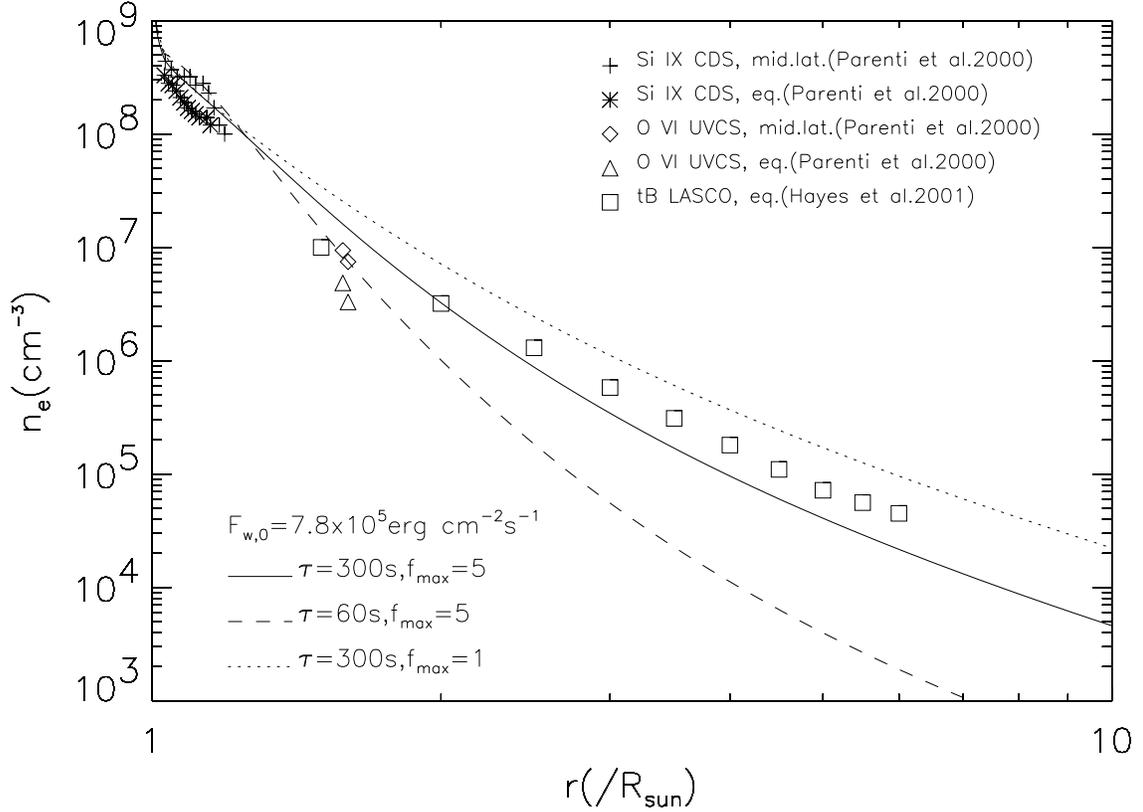} 
\caption{Comparison of electron density distributions. Lines are model results 
and points are observational data. Models adopt the same 
$F_{\rm w,0}= 7.8\times 10^5$erg cm$^{-2}$s$^{-1}$ and different sets of  
$(\tau({\rm s}), f_{\rm max})=(300,5)$ (solid), (60,5) (dashed), and 
(300,1) (dotted). Pluses are data from CDS/SOHO observation of the 
mid-latitude streamer, asterisks from CDS/SOHO observation of the 
equatorial streamer, diamonds from UVCS/SOHO observation of the 
mid-latitude streamer, and triangles from UVCS/SOHO observation of the 
equatorial streamer \citep{pbp00}. 
Squares are derived from observation of total brightness by 
LASCO/SOHO \citep{hvh01}. 
}
\label{fig:obsne}
\end{figure}

In fig.\ref{fig:obsne}, we display our results of electron density with 
observation of the streamer \citep{pbp00,hvh01}. 
Our models adopt an identical $F_{\rm w,0} = 7.8\times 10^5$erg 
cm$^{-2}$s$^{-1}$ and three sets of   
$(\tau({\rm s}), f_{\rm max})=(300, 5), (60,5), (300, 1)$ 
(same as fig.\ref{fig:taudp}). 
As for the observation, 
we show results derived from a line ratio of Si IX by CDS/SOHO for a region of 
$1.02 - 1.19 R_{\odot}$,  a ratio of radiative and collisional intensities 
of O VI line by UVCS/SOHO for a region of $1.58 - 1.6R_{\odot}$ 
\citep{pbp00}, and total brightness obtained from LASCO/SOHO for a region 
$1.5 R_{\odot} \lesssim r \lesssim 6 R_{\odot}$ \citep{hvh01}. 
With respect to the CDS and UVCS data, both results for equatorial and 
mid-latitude streamer regions are displayed.  

Our results, adopting the same  $F_{\rm w,0}$, 
show almost identical density in very inner part, being independent 
of $\tau$ and $f_{\rm max}$ (\S \ref{sc:tmxptr} and fig.\ref{fig:fwptr}). 
The figure exhibits that the adopted value, 
$F_{\rm w,0}=7.8 \times 10^5$erg cm$^{-2}$s$^{-1}$, reproduces the 
CDS observation of the mid-latitude streamer well.  To fit to the data of the 
equatorial streamer, slightly smaller input of $F_{\rm w,0}$ is favored. 
In our model, $\tau$ and $f_{\rm max}$ control slope of density 
distribution. $\tau$ influences temperature in an intermediate 
region of $2R_{\odot} \lesssim r \lesssim 10R_{\odot}$ 
(fig.\ref{fig:taudp}), hence, regulates the density 
scale height (or decreasing slope of density) in that region. The data 
based on LASCO/SOHO indicates that models considering larger $\tau$(=300s) 
are more likely. $f_{\rm max}$ determines density decrease in a region of 
$1 \sim  2 R_{\odot}$. The data of CDS ($\lesssim 1.2 
R_{\odot}$) and UVCS ($\sim 1.6 R_{\odot}$) exhibits the drastic decrease 
of density, which indicates that non-radial expansion is desired. 
Adjustment of the other parameters of the flow tube geometry 
($r_1$ and $\sigma$; eq.(\ref{eq:ftg})) would give the still better fit.   
Although we do not further search the best parameter set to fit to the 
observation, the figure indicates that our model could reproduce the 
observed density profile by the choice of the appropriate 
parameters ($F_{\rm w,0}\simeq (5-8)\times 10^5$erg cm$^{-2}$s$^{-1}$, 
$\tau\approx 300$s, and $f_{\rm max}\approx 5$).

\subsubsection{Temperature Distribution}
Figure \ref{fig:obste} compares our results of temperature distribution with 
observation in the streamer region. Our models employ the same parameter sets 
as in fig.\ref{fig:obsne}, and the data were electron temperature obtained 
from observations of a line ratio of Fe XIII/Fe X by the CDS and UVCS/SOHO 
\citep{pbp00}. 
Although the observed data are electron temperature, they are supposed 
to represent the plasma temperature because electron-ion equilibrium 
is attained in dense streamer region of $r\lesssim 2R_{\odot}$ \citep{rskn98}. 
Therefore, it is reasonable to compare them to the results of our model 
considering one-fluid coronal plasma. 
A case employing 
$(F_{\rm w,0}, \tau, f_{\rm max})=(7.8\times 10^5, 300, 5)$ gives reasonable 
peak temperature of $\simeq 1.5 \times 10^6$K. However, none of our models 
can reproduce the observed location of $T_{\rm max}$. 
While the location is observationally inferred to be between $1.2 R_{\odot}$ 
and $1.6 R_{\odot}$, all of our models give $r_{T{\rm max}}< 1.2 R_{\odot}$. 
This is because the dissipation length of the N-waves is essentially 
short, even though one considers long period-waves that are generated in the 
corona.  We can summarize that acoustic waves excited in the corona could 
certainly heat the surrounding plasma to $T>10^6$K, 
however, they cannot maintain the high temperature 
till the sufficiently distant region by themselves. Therefore, the 
cooperations with other heating sources with larger dissipation length 
are necessary to explain the observed solar corona.

\subsubsection{Velocity Distribution}
In fig.\ref{fig:obsvl}, we show the results of velocity distribution of the 
solar wind, with observational results in the low-latitude 
streamer (shaded). 
The observational data are from \citet{she97}, who determined velocity 
profile between 2 and 30 $R_{\odot}$ from measurements of about 65 moving 
objects in the streamer belt.
They used two different technic in deriving the results, whereas the  
shaded area displayed in fig.\ref{fig:obsvl} are based on the straight-line 
fit method (an upper panel of fig.6 in \citet{she97}; The shaded region is 
traced from that figure). 
Figure \ref{fig:obsvl} indicates that the resultant velocity of our models 
is lower than the observed 
data in the whole region. The observation exhibits rapid 
acceleration in $3 \sim 5R_{\odot}$, while our results show gradual 
acceleration by thermal gas pressure, since the N-waves are damped in 
$<1.5 R_{\odot}$ and wave pressure cannot contribute to the acceleration of 
the wind flow, as shown in figs.\ref{fig:taudp} and \ref{fig:fwdp}. 
Other mechanisms are also required in acceleration of the 
low-speed wind.

\section{Summary and Discussions}
\label{sc:smds}
We have investigated the process of acoustic waves generated in the corona 
as a heating source especially in the mid- to low-latitude corona where the 
low-speed winds come from. 
We have found that the acoustic waves with $\tau\ge 60$s and $F_{\rm w,0}
\ge 2\times 10^5$erg cm$^{-2}$s$^{-1}$ could certainly heat up the 
ambient plasma to $T\ge 10^6$K by the dissipation of the N-waves even in a 
flow tube with the areal expansion of $f_{\rm max}=5$, 
by balancing with the losses of the 
radiative cooling, the downward thermal conduction, and the adiabatic loss by 
the solar wind. 
Due to its dissipative character, the dissipation of N-waves effectively works 
in the inner corona and reproduces the density profile observed in the 
streamer region. However, it cannot contribute to the heating of the 
outer corona, since most of the wave 
energy is damped within a region of few-tenth of the solar radius even if 
one considers 
the waves with a long period of $\tau=300$s. As a result, it is impossible 
to explain the observed temperature profile and flow velocity of the 
low-speed wind only by that process. 
Therefore, other mechanisms with the larger dissipation length should play 
a role cooperatively in the coronal 
heating and the acceleration of the low-speed wind.

\begin{figure}
\figurenum{9} 
\epsscale{1} 
\plotone{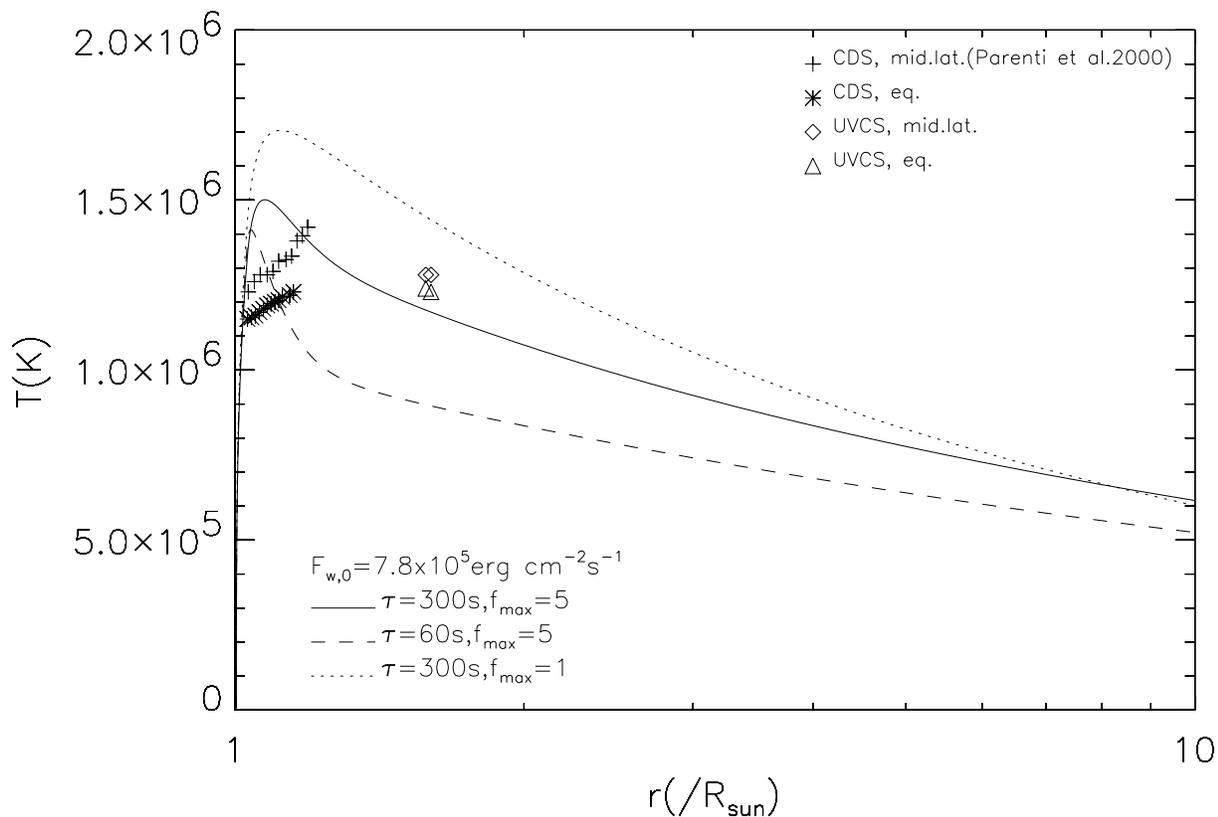} 
\caption{Comparison of electron density distributions. Respective lines 
represent results of the models adopting the same parameter sets as in 
fig.\ref{fig:obsne}. Pluses are data from CDS/SOHO observation of the 
mid-latitude streamer, asterisks from CDS/SOHO observation of the 
equatorial streamer, diamonds from UVCS/SOHO observation of the 
mid-latitude streamer, and triangles from UVCS/SOHO observation of the 
equatorial streamer \citep{pbp00}. 
}
\label{fig:obste}
\end{figure}

\begin{figure}
\figurenum{10} 
\epsscale{1} 
\plotone{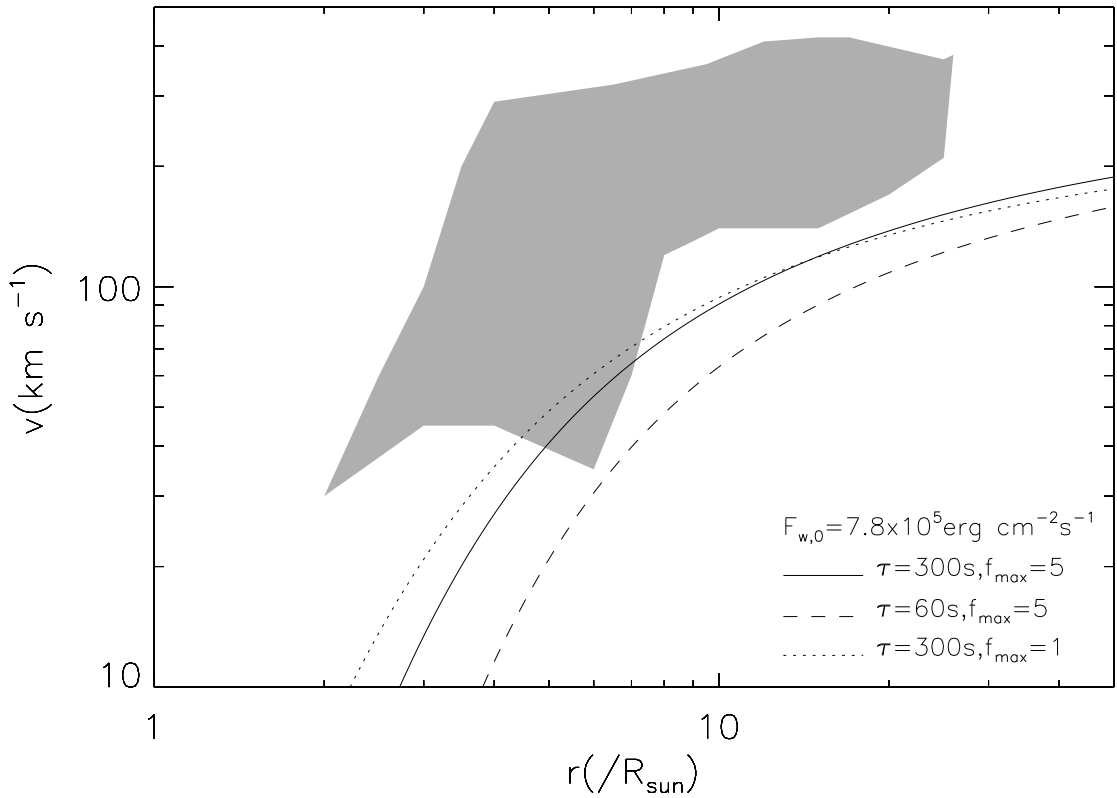} 
\caption{Comparison of velocity profile of the solar wind. Respective lines 
represent results of the models adopting the same parameter sets as in 
fig.\ref{fig:obsne}. Shaded region is observational data derived from 
measurements of about 65 moving objects in the streamer belt \citep{she97} 
(See text).
}
 \label{fig:obsvl}
 \end{figure}

Small-scale reconnection events and spicules also predict  
generation of fast shock waves \citep{lw00,lee01,hol82}, though we have 
concentrated on the role of the N-waves (slow shocks along the magnetic 
field line) excited by those events in this paper.
The dissipation length of the fast shocks must be larger than N-waves with 
the identical period, since 
their phase speed is $\gtrsim v_{\rm A}$, which is much larger than that of 
the N-waves ($\sim c_{\rm s}$) for low-$\beta$ coronal plasma.  
Hence, cooperation with the fast shocks would give a more distant location 
of $T_{\rm max}$ to match the observed temperature profile 
(fig.\ref{fig:obste}), and would let the 
acceleration of the wind flow continue in the outer corona, as suggested 
by observation of the streamer belt (fig.\ref{fig:obsvl}).  

Finally, we had better remark an issue on anisotropic and selective heating 
of ions. According to recent observations \citep{ssp02}, 
O VI ions in the streamer have high perpendicular kinetic temperature, 
though this is not so extreme as that observed in the high-speed wind. 
Our results show that the deposition of energy 
and momentum from the N-waves to the ambient gas is completed in the region of 
$r\lesssim 1.5 R_{\odot}$, where the particles are thermally well coupled. 
As a result, whatever anisotropies obtained by the N-wave heating 
would be wiped out and all the ions would have 
isotropic kinetic temperature. The process of the N-waves cannot explain the 
observed anisotropies, which also indicates that the other mechanisms 
causing the anisotropic heating have to operate simultaneously.

We thank K. Shibata, K. Ohki, K. Omukai, K. Tomisaka, H. Saio, S. Nitta, 
S. Inutsuka, T. Kudoh, Y. Yoshii, T. Kajino, and A. Tohsaki 
as well as members of DTAP in NAOJ for 
many valuable and critical comments and Y. Mclean for improvement of 
presentation in this paper. 
The author is supported by the JSPS Research Fellowship for Young Scientists,
grant 5936.

\begin{thebibliography}{}

\bibitem[Allen(1973)]{aln73}
Allen, C. W. 1973, Astrophysical Quantities(London:Athlone) 

\bibitem[Aschwanden, Poland, \& Rabin (2001)]{apr01}
Aschwanden, M. J., Poland, A. I., \& Rabin, D. M. 2001, \araa, 39, 175

\bibitem[Bastian, Benz, \& Gary (1998)]{bbg98}
Bastian, T. S., Benz, A. O., \& Gary, D. E. 1998, \araa, 36, 131
 
%

\bibitem[Cranmer, Field, \& Kohl(1999)]{cfk99}
Cranmer, S. R., Field, G. B., \& Kohl, J. L. 1999, \apj, 518, 937

\bibitem[Erd\'{e}lyi et al.(1998)]{edpw98}
Erd\'{e}lyi, R., Doyle, J. G., Perez, M. E., \& Wilhelm, K. 1998, 
\aap, 337, 287

\bibitem[Furusawa \& Sakai(2000)]{fs00}
Furusawa, K. \& Sakai, J. I. 2000, \apj, 540, 1156

\bibitem[Habbal et al.(1997)]{hwf97}
Habbal, S. R., Woo, R., Fineschi, S., O'neal, R., Kohl, J., Noci, G., 
\& Korendyke, C. 1997, \apjl, 489, L103

\bibitem[Hammer(1982a)]{hm82a}
Hammer, R. 1982a, \apj, 259, 767

\bibitem[Hammer(1982b)]{hm82b}
Hammer, R. 1982b, \apj, 259, 779

\bibitem[Hassler et al.(1990)]{hrsh90}
Hassler, D. M., Rottman, G. J., Shoub, E. C., \& Holzer, T. E. 1990, \apjl, 
348, L77

\bibitem[Hayes, Vourlidas, \& Howard(2001)]{hvh01}
Hayes, A. P., Vourlidas, A., \& Howard, R. A. 2001, \apj, 548, 1081

\bibitem[Hollweg(1982)]{hol82}
Hollweg, J. V. 1982, \apj, 254, 806 

\bibitem[Hollweg(1992)]{hol92}
Hollweg, J. V. 1992, \apj, 389, 731 

\bibitem[Hollweg(1999)]{hol99}
Hollweg, J. V. 1999, \jgr, 104, 24781 

\bibitem[Hollweg, Jackson, \& Galloway (1982)]{hjg82}
Hollweg, J. V., Jackson, S., \& Galloway, D. 1982, Sol.Phys., 75, 35

\bibitem[Hudson(1991)]{hd91}
Hudson, H. S. 1991, Sol.Phys., 133, 357

\bibitem[Kopp \& Orrall(1976)]{ko76}
Kopp, R. A. \& Orall, F. Q. 1976, \aap, 53,363

\bibitem[Krucker \& Benz(1998)]{kb98}
Krucker, S. \& Benz, A. O. 1998, \apjl, 501, L213

\bibitem[Kudoh \& Shibata(1999)]{ks99}
Kudoh, T. \& Shibata, K. 1999, \apj, 514, 493

\bibitem[Landau \& Lifshitz(1959)]{LL59}
Landau, L. D. \& Lifshitz, E. M. 1959, 'Fluid Mechanics'  

\bibitem[Landini \& Monsignori-Fossi(1990)]{LM90}
Landini, M. \& Monsignori-Fossi, B. C. 1990, \aaps, 82, 229

\bibitem[Lee \& Wu(2000)]{lw00}
Lee, L. C. \& Wu, B. H. 2000, \apj, 535, 1014

\bibitem[Lee(2001)]{lee01}
Lee, L. C.  2001, \ssr, 95, 95

\bibitem[McWhirter, Thonemann, \& Wilson(1975)]{mtw75}
McWhirter, R. W. P., Thonemann, P. C., \& Wilson, R. 1975, \aap, 40, 63

\bibitem[Mihalas \& Mihalas(1984)]{mm84}
Mihalas, D. \& Mihalas, B. W. 1984, 'Foundation of Radiation Hydrodynamics', 
Oxford Unuversity Press

\bibitem[Nishio et al. (1997)]{nso97}
Nishio, M., Yaji, K., Kosugi, T., Nakajima, H., \& Sakurai, T. 1997, \apj, 
489, 976

\bibitem[Osterbrock (1961)]{ost61}
Osterbrock, D. E. 1961, \apj, 134, 347

\bibitem[Parenti et al.(2000)]{pbp00}
Parenti, S., Bromage, B. J. I., Poletto, G., Noci, G., Raymond, J. C., \& 
Bromage, G. E. 2000, \aap, 363, 800

\bibitem[Parker(1958)]{pkr58}
Parker, E. N. 1958, \apj, 128, 664


\bibitem[Parnell \& Jupp(2000)]{pj00}
Parnell, C. E. \& Jupp, P. E. 2000, \apj, 529, 554

\bibitem[Raymond et al.(1998)]{rskn98}
Raymond, J. C., Suleiman, R., Kohl, J. L., \& Noci, G. 1998, \ssr, 85, 283

\bibitem[Roald, Sturrock, \& Wolfson(2000)]{rsw00}
Roald, C. B., Sturrock, P. A., \& Wolfson, R.  2000, \apj, 538, 960 (RSW)

\bibitem[Rosner \& Vaiana(1977)]{rv77}
Rosner, R. \& Vaiana, G. S. 1977, \apj, 216, 141

\bibitem[Rosner, Tucker, \& Vaiana(1978)]{rtv78}
Rosner, R., Tucker, W. H., \& Vaiana, G. S. 1978, \apj, 220, 643

\bibitem[Sakai et al.(2000)]{sky00}
Sakai, J.I., Kawata, T., Yoshida, K., Furusawa, K., \& Cramer, N. F. 2000,
\apj, 537, 1063

\bibitem[Sandb{\ae}k \& Leer(1994)]{sl94}
Sandb{\ae}k, \O. \& Leer, E. 1994, \apj, 423, 500

\bibitem[Sheeley et al.(1997)]{she97}
Sheeley, N. R. Jr. et al. 1997, \apj, 484, 472

\bibitem[Stein \& Schwartz(1972)]{ss72}
Stein, R. F. \& Schwartz, R. A. 1972, \apj, 177, 807 (SS)

\bibitem[Strachan et al.(2002)]{ssp02}
Strachan, L, Suleiman, R., Panasyuk, A. V., Biesecker, D. A., \& Kohl, J. L. 
2002, \apj, 571, 1008

\bibitem[Sturrock(1999)]{str99}
Sturrock, P. A.  1999, \apj, 521, 451 (S99)

\bibitem[Sturrock, Roald, \& Wolfson(2000)]{srw00}
Sturrock, P. A., Roald, C. B., \& Wolfson, R.  1999, \apjl, 524, L75

\bibitem[Tarbell, Ryutova, \& Covington(1999)]{trc99}
Tarbell, T., Ryutova, M., \& Covington, J. 1999, \apjl, 514, L47

\bibitem[Tsuneta et al.(1992)]{tnt92}
Tsuneta, S. et al. 1992, \pasj, 44, L63

\bibitem[Tsuneta (1996)]{tnt96}
Tsuneta, S. 1996, \apj, 456, 840

\bibitem[Ulmschneider(1971)]{ulm71}
Ulmschneider, P. 1971, \aap, 12, 297

\bibitem[Ulrich(1996)]{ulr96}
Ulrich, R. K. 1996, \apj, 465, 436

\bibitem[Withbroe \& Noyes(1977)]{wn77}
Withbroe, G. L. \& Noyes, R. W. 1977, \araa, 15, 363

\bibitem[Withbroe(1988)]{wtb88}
Withbroe, G. L. 1988, \apj, 325, 442
\end{thebibliography}
\end{document}